\documentclass[journal=jctc,manuscript=article,layout=two-column]{achemso}
\setkeys{acs}{articletitle = true}
\setkeys{acs}{maxauthors=10}
\setkeys{acs}{etalmode=truncate}

\makeatletter
\let\l@addto@macro\relax
\makeatother
\usepackage[fontsize=10pt]{scrextend}
\SectionNumbersOn

\usepackage[version=4]{mhchem} 
\usepackage{float}
\usepackage{graphicx}
\usepackage{amssymb}
\usepackage{lineno}
\usepackage{xcolor}
\usepackage{amsmath}
\usepackage{booktabs}
\usepackage{multirow}
\usepackage{lipsum}
\usepackage{color}
\usepackage{soul}
\usepackage{booktabs}
\usepackage[english]{babel}

\usepackage{hyperref}
\usepackage{amsmath}

\author{Amanda D. Torres}
\email{dornelaamanda@pos.iq.ufrj.br}
\affiliation[Universidade Federal do Rio de Janeiro]
{Chemistry Institute, Federal University of Rio de Janeiro, Rio de Janeiro, Brazil - 21941-909}

\author{Bidhan Chandra Garain}
\affiliation[University College London]{Department of Chemistry, University College London, 20 Gordon St., London WC1H 0AJ, UK}

\author{Mario Barbatti}
\affiliation[Aix-Marseille Université]{Aix-Marseille University, CNRS, ICR, 13397 Marseille, France}
\alsoaffiliation[Institut Universitaire de France]{Institut Universitaire de France, 75231 Paris, France}

\author{Ricardo R. Oliveira}
\email{rrodrigues.iq@gmail.com}
\affiliation[Universidade Federal do Rio de Janeiro]
{Chemistry Institute, Federal University of Rio de Janeiro, Rio de Janeiro, Brazil - 21941-909}

\author{Alexandre B. Rocha}
\email{rocha@iq.ufrj.br}
\affiliation[Universidade Federal do Rio de Janeiro]
{Chemistry Institute, Federal University of Rio de Janeiro, Rio de Janeiro, Brazil - 21941-909}

\title{VCMC: A Python Monte Carlo Workflow for Optical Electron-Phonon Coupling on Crystalline Systems}

\begin{document}

\renewcommand\tocentryname{TOC Graphic}
\begin{tocentry}
\centering
\includegraphics[width=4.5cm]{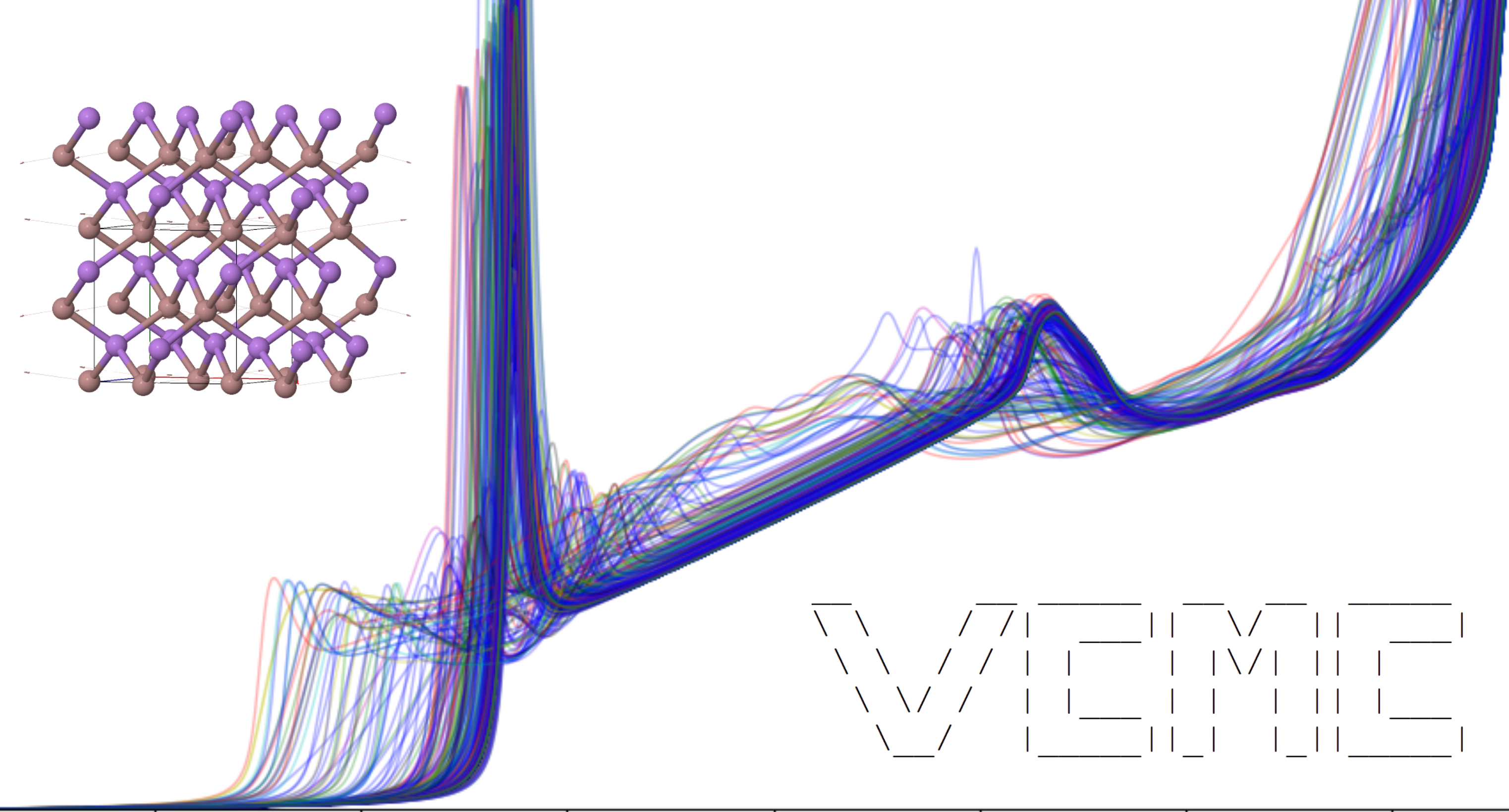}
\end{tocentry}

\maketitle

\begin{abstract}
Vibronic or electron-phonon coupling effects in optical spectra play a central role in the spectroscopy of crystalline materials, but their explicit treatment remains computationally demanding because of the large number of phonon degrees of freedom. Here, we present VCMC, a Monte Carlo protocol and Python workflow for simulating vibronic contributions to optical spectra of crystalline systems. Nuclear configurations are sampled from harmonic vibrational distributions around the equilibrium structure, and the optical response is obtained by averaging the frequency-dependent dielectric function over the resulting ensemble of distorted geometries. To reduce the number of electronic-structure calculations, the workflow combines structural descriptors, dimensionality reduction, and clustering to select representative configurations from the Monte Carlo ensemble. The method is tested on Si, GaAs and crystalline formaldehyde. For Si, GaAs, and formaldehyde, VCMC reproduces vibronically induced spectral features and broadening effects that are absent or poorly described at the equilibrium geometry. These results show that ensemble-based vibronic sampling, combined with clustering, provides a practical route for including nuclear-motion effects in optical spectra of periodic systems.
\end{abstract}

\section*{Keywords}
Vibronic Coupling, Crystals, Monte Carlo, Unsupervised Learning

\section{Introduction}

Electron-phonon coupling effects, also known as vibronic coupling, are known to play a crucial role in a wide range of physical phenomena in solids \cite{Giustino2017}, including superconductivity \cite{Liu1996}, electrical and thermal transport \cite{Quan2021}, electron scattering processes \cite{Liu1996}, and the temperature dependence of the electron bands \cite{Giustino2017}. Despite the field's importance, for many years it relied on semi-empirical models, and it was only after the implementation of the Density Functional Perturbation Theory (DFPT) that its development from first principles began to grow \cite{Giustino2017}. In recent years, considerable effort has been devoted to calculate electron-phonon matrix elements using a variety of approaches \cite{Backman2022, Li2024, Garai-Marin2026, Giustino2007, Lee2023, Ye2023, Ponce2016, Radevych2025, Poliukhin2026, Ponce2025, Houtput2025}.

Considering the optical response of the electron-phonon coupling, one of the earliest and most influential theories regarding electronic transitions is that of Hall, Bardeen, and Blatt (HBB) for indirect gap absorption. This theory states that electrons experience a time-dependent potential arising from the vibrations of the atoms around them \cite{Hall1954}. This framework later served as the basis for first-principles calculations of phonon-assisted absorption of \ce{Si} \cite{Noffsinger2012}, \ce{GaN} \cite{Kioupakis2010}, and \ce{SnO_2} \cite{Peelaers2015}. Zacharias and Giustino \cite{Zacharias2015} also analyzed the problem of indirect gap absorption, but using a stochastic approach. Using a Williams-Lax-based \cite{Lax1952} dielectric function, they performed Monte Carlo sampling of atomic configurations and averaged over all configurations. This is a similar approach to the one done in this work. 

The importance of nuclear motion in electronic spectroscopy has long been recognized through the Franck-Condon principle \cite{Zuehlsdorff2019}. Within the Born-Oppenheimer separation, a molecular state is written as a product of electronic and nuclear states. The Franck-Condon approximation further assumes that the electronic transition dipole moment varies only weakly with the nuclear coordinates over the region sampled by the vibrational states. Under this assumption, the transition amplitude factorizes into an electronic and a nuclear contribution. For light polarized along the unit vector $\mathbf{e}$, one obtains

\begin{equation}
M_{fi}=\langle \psi_f | \hat{\boldsymbol{\mu}}\cdot\mathbf{e} | \psi_i \rangle \langle \chi_f | \chi_i\rangle ,
\end{equation}

\noindent and the corresponding transition intensity is proportional to

\begin{equation}
I_{fi}\propto\left|\langle \psi_f | \hat{\boldsymbol{\mu}}\cdot\mathbf{e} | \psi_i \rangle\right|^2\left|\langle \chi_f | \chi_i \rangle\right|^2 .
\end{equation}

Here, $\psi_i$ and $\psi_f$ denote the initial and final electronic states, while $\chi_i$ and $\chi_f$ denote the corresponding vibrational states. The nuclear term, usually called the Franck-Condon factor, measures the vibrational overlap and modulates the electronic transition strength.

A more general and widely used treatment of vibronic effects relaxes the Condon approximation by explicitly accounting for the dependence of the transition dipole moment on nuclear coordinates. This leads to the Herzberg–Teller expansion \cite{HerzbergTeller1933}, in which the dipole moment is expressed as a power series in the normal-mode coordinates, and the total transition probability is obtained by summing over vibrational states. 

Within this framework, direct methods \cite{Rocha2000} offer a simple and physically transparent way to include vibronic effects: the potential energy surface is obtained along selected normal modes, and the electronic response is evaluated at the resulting displaced geometries. In molecular spectroscopy, these methods and related approximations have been extensively tested, with different choices yielding varying balances between accuracy and computational cost \cite{Torres2022, Cerdan2022, Feher2021, Noffsinger2012, Zacharias2015}. Comparisons of systems such as formaldehyde have helped expose the practical strengths and limitations of each treatment. The situation is less mature for periodic systems, where the same vibronic problem becomes substantially more demanding \cite{Grochala2003, Kundu2023, Zacharias2020}.

In solids, the vibrational degrees of freedom are described by phonon modes characterized by a wave vector \cite{Giustino2017}, resulting in a large number of vibrational modes even for relatively small unit cells. For each wave vector, three acoustic and $3N-3$ optical modes are present, where $N$ is the number of atoms in the primitive cell. The number of vibrational modes, combined with their collective and often delocalized nature, leads to a dramatic increase in computational cost when explicitly accounting for vibronic effects using approaches originally developed for molecular systems. As a consequence, mode-by-mode treatments become impractical in periodic systems. In particular, extending vibronic concepts commonly used in molecular spectroscopy to periodic systems in a manner that is computationally feasible and physically accurate remains an open problem.

Understanding and predicting detailed electronic spectra is essential for the design of new materials and for extracting physical insights into solid-state systems. Despite significant advances in the first-principles description of electron-phonon interactions and vibronic effects, a conceptual gap remains between approaches traditionally employed in solid-state physics and those developed in molecular spectroscopy. Methods based on DFPT or frozen-phonon schemes efficiently describe phonon-induced renormalization of electronic properties but do not directly yield spectroscopic observables in a form comparable to molecular vibronic theories. Conversely, vibronic approaches widely used in chemistry provide an intuitive and accurate description of spectral line shapes but become computationally prohibitive in periodic systems due to the large number of phonon modes and wave vectors involved. In this context, ensemble-based approaches offer a natural route to reformulate vibronic effects in solids in terms of statistical sampling of nuclear configurations. 

In this work, we address this problem by developing a Monte Carlo protocol to include vibronic effects within the Nuclear Ensemble approach \cite{Karsai2018, Crespo-Otero2012, Barbatti2010}. Nuclear configurations are sampled from a harmonic distribution around the equilibrium geometry, and the imaginary part of the frequency-dependent dielectric function is evaluated for each distorted structure. The final spectrum is then obtained by averaging over the ensemble. To reduce computational cost while retaining the relevant spectral information, we further combine this sampling strategy with a clustering procedure, as described in the next section. This offers automated selection of representative structures, enabling the calculation of spectroscopic properties at a feasible computational cost for periodic crystalline systems. 


\section{Theoretical Approach}

Within the Born-Oppenheimer approximation, a vibronic state is written as a product of an electronic state and a vibrational state. In the electric-dipole approximation, the interaction between matter and the electromagnetic field is mediated by the molecular transition dipole moment. From this quantity, one can derive several spectroscopic observables, including oscillator strengths, absorption cross sections, Einstein coefficients, and the imaginary part of the frequency-dependent dielectric function.

In the Condon approximation, the electronic transition dipole moment is assumed to be independent of the nuclear coordinates, so that the vibronic transition amplitude separates into an electronic and a vibrational contribution. However, when this approximation breaks down, the nuclear dependence of the transition dipole moment must be retained. This situation is particularly important for vibronically allowed transitions, where nuclear motion can give intensity to an otherwise electronically forbidden transition. For a transition from the vibronic state $(i,\nu)$ to $(f,\nu')$, the intensity may then be written as

\begin{equation}
I_{i\nu \rightarrow f\nu'}\propto\sum_{\alpha=x,y,z}\left|\int\chi_{f\nu'}^{*}(\mathbf Q)M_{if}^{\alpha}(\mathbf Q)\chi_{i\nu}(\mathbf Q) d\mathbf Q\right|^2 
\end{equation}

where

\begin{equation}
M_{fi}^{\alpha}(\mathbf Q)=\left\langle\psi_f(\mathbf r;\mathbf Q)\middle|\hat{\mu}_{\alpha}\middle|\psi_i(\mathbf r;\mathbf Q)\right\rangle.
\end{equation}

Here, $\psi_i(\mathbf r;\mathbf Q)$ and $\psi_{f}(\mathbf r;\mathbf Q)$ are electronic states that depend parametrically on the nuclear geometry $\mathbf Q$, while $\chi_{i\nu}(\mathbf Q)$ and $\chi_{f\nu'}(\mathbf Q)$ are vibrational states associated with the corresponding electronic surfaces. The vector $\mathbf r$ denotes the electronic coordinates, $\mathbf {Q} $ the normal-mode coordinates, and $\hat{\mu}_{\alpha}$ the $\alpha$ Cartesian component of the electric dipole moment operator. Depending on the form adopted for $M_{if}^{\alpha}(\mathbf Q)$, for example, through a Taylor expansion around the equilibrium geometry, the vibrational integrals can be evaluated analytically for harmonic normal modes.  

A practical alternative is to replace the explicit multidimensional vibronic integral by Monte Carlo sampling over nuclear configurations. Rather than evaluating the response over the full vibrational coordinate space, nuclear geometries are drawn from the relevant vibrational probability distribution, and the spectrum is estimated as an ensemble average. This strategy is particularly attractive for periodic systems, where the large number of phonon modes, electronic bands, and near-degenerate states makes an explicit mode-by-mode analytical treatment impractical. In this context, Monte Carlo sampling provides a direct way to include thermal and zero-point nuclear fluctuations in the optical response, at the price of requiring a sufficiently large ensemble to achieve statistical convergence.

To achieve this goal, we developed Vibronic Coupling from Monte Carlo (VCMC), a Python workflow for simulating vibronic effects in optical spectra through Monte Carlo sampling of nuclear configurations. Starting from phonon calculations, the program samples displacements from harmonic vibrational distributions and generates the corresponding distorted geometries, from which the optical response is evaluated for each sampled structure. To reduce the number of electronic-structure calculations required for ensemble averaging, the workflow further includes a clustering strategy based on structural descriptors and dimensionality reduction, enabling the selection of representative geometries from the Monte Carlo ensemble. In its current implementation, VCMC is designed to be used with VASP.

In the sampling process, VCMC generates new geometries by randomly displacing the normal-mode coordinates. For each mode, $j$, the program calculates the Boltzmann probabilities of the vibrational states at the selected temperature. During the generation of each geometry, it randomly selects a vibrational state, $\nu$, for each mode according to this distribution. The state distribution is truncated so that the discarded Boltzmann tail is below 0.0001\%. The probability of selecting vibrational state \(v\) for mode \(j\) is given by

\begin{equation}
P_j(v)= \frac{(1-q_j)q_j^v}{\displaystyle\sum_{v'=0}^{v_{\max}}(1-q_j)q_j^{v'}}, \qquad v=0,\ldots,v_{\max},
\end{equation}

where

\begin{equation}
q_j=\exp\!\left(-\frac{\hbar\omega_j}{k_{\mathrm B}T}\right),
\end{equation}

and \(\omega_j\) is the angular frequency of mode \(j\), \(T\) is the temperature, and \(v_{\max,j}\) is the highest vibrational state retained for that mode.

After selecting a vibrational state for a mode, the program samples a candidate normal-mode coordinate, \(Q_f\). A random number \(x_{\mathrm{random}}\) is selected from the range \([0,1]\), and the candidate coordinate is generated as

\begin{equation}
Q_f = (2*x_{random}*Q_{max}) - Q_{max}
\end{equation}

\noindent in a way that if $x_{random} = 0$, then $Q_f = -Q_{max}$, and if $x_{random} = 1$, then $Q_f =  Q_{max}$. This condition ensures that we remain within a finite sampling window. The maximum displacement, $Q_{\max,jv}$, is defined using the standard deviation of the harmonic-oscillator probability density for each mode $j$ and vibrational state $v$:

\begin{equation}
\sigma_{jv}=\sqrt{\frac{v+\tfrac{1}{2}}{\alpha_j}}
\qquad
\alpha_j=\frac{\omega_j}{\hbar}.   
\end{equation}

Here, $Q_j$ is a mass-weighted normal coordinate, $\omega_j$ is the angular frequency of mode $j$, and all quantities must be expressed in compatible units. In the implementation, frequencies are expressed in rad s$^{-1}$ and normal coordinates in $\sqrt{\mathrm{u}}\,\text{\AA}$, with the corresponding unit conversion included when evaluating $\alpha_j$.

The sampling interval is defined as

\begin{equation}
Q_{\max,jv}=c\,\sigma_{jv},
\end{equation}

where $c$ is a user-defined cutoff, set to 6 by default. This finite interval truncates the tails of the probability density. The probability retained within the interval depends on the vibrational state and the selected cutoff.

For each candidate displacement $Q_f$, the harmonic probability density is evaluated as

\begin{equation}
\rho_{jv}(Q_f)=|\psi_{jv}(Q_f)|^2,
\end{equation}

where

\begin{equation}
\psi_{jv}(Q)=\frac{\alpha_j^{1/4}}{\pi^{1/4}\sqrt{2^v v!}}H_v\!\left(\sqrt{\alpha_j}\,Q\right)\exp\!\left(-\frac{\alpha_j Q^2}{2}\right).
\end{equation}

Here, $H_v$ is the Hermite polynomial of order $v$. A random value $y_{\mathrm{random}}$ is then selected from $[0,\rho_{jv,\max}]$, where

\begin{equation}
\rho_{jv,\max}=\max_{Q\in[-Q_{\max},Q_{\max}]}\rho_{jv}(Q).
\end{equation}

\begin{figure}
    \centering
    \includegraphics[width=0.75\linewidth]{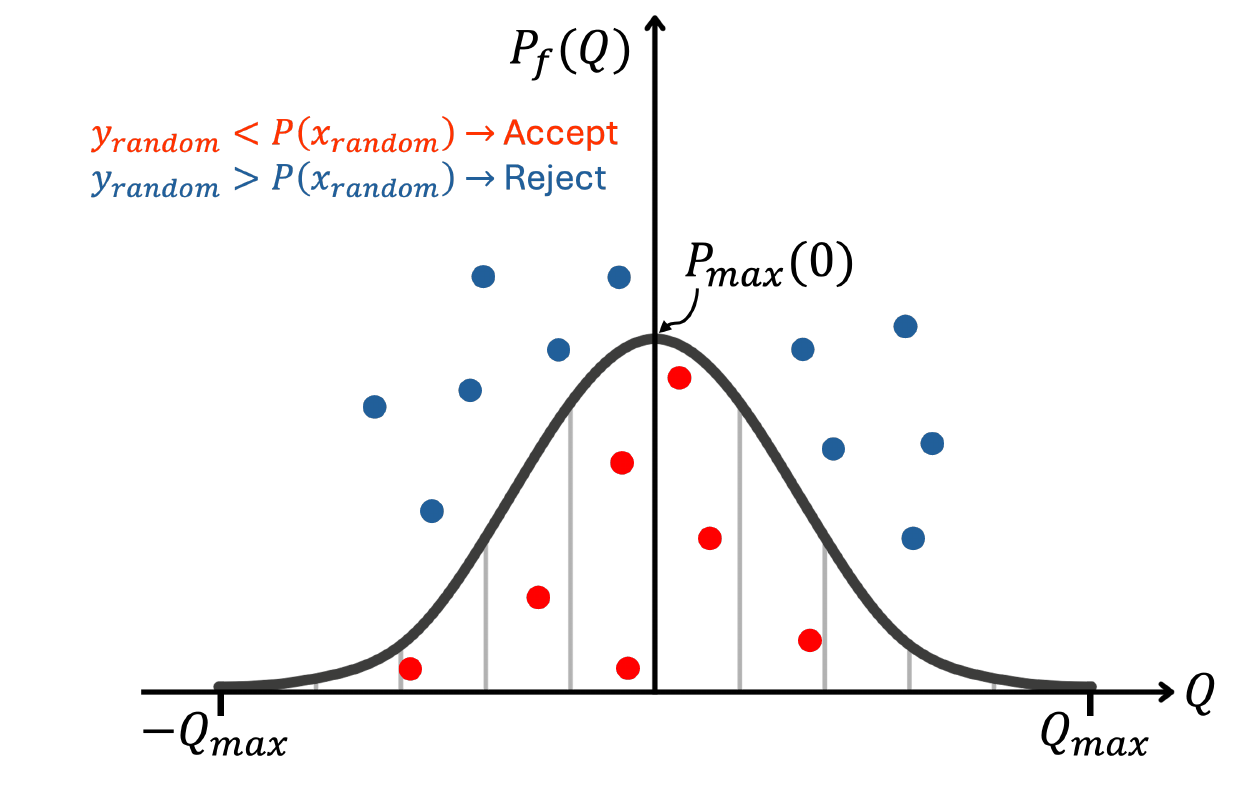}
    \caption{A schematic representation of the Monte Carlo sampling process, illustrating accepted and rejected points on a Gaussian distribution that depicts the harmonic probability function.}
    \label{fig:MonteCarlo}
\end{figure}

The candidate displacement is accepted if $y_{\mathrm{random}}\leq \rho_{jv}(Q_f)$. Otherwise, it is rejected and a new candidate displacement is generated. This rejection-sampling procedure makes the accepted displacements follow the harmonic probability density within the selected sampling interval. An example is shown in Figure~\ref{fig:MonteCarlo}.

\begin{figure}
    \centering
    \includegraphics[width=0.75\linewidth]{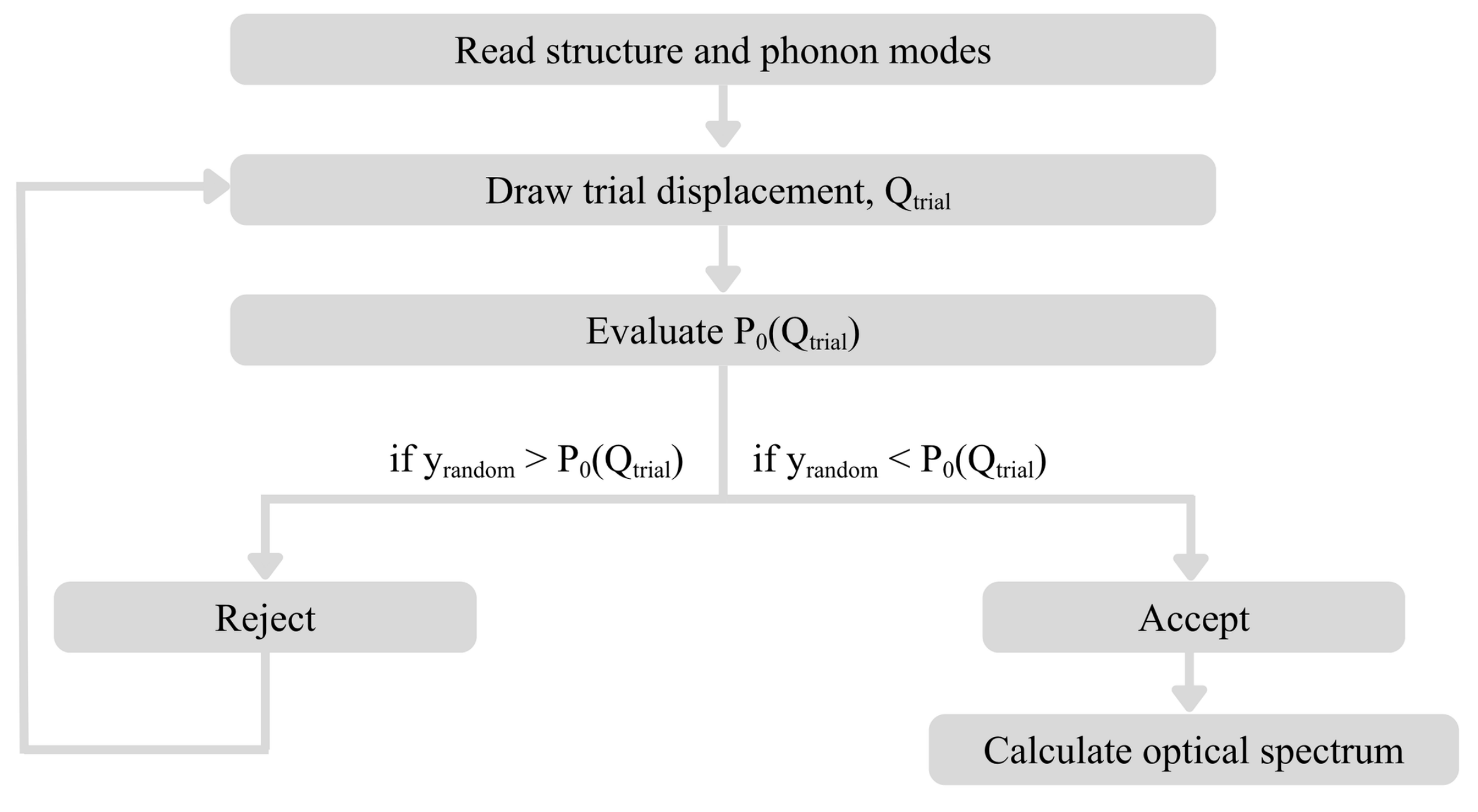}
    \caption{Workflow for generating distorted geometries by rejection sampling from the harmonic vibrational probability distribution. The accepted displacements are transformed into Cartesian coordinates and used for optical-spectrum calculations.}
    \label{fig:workflow1}
\end{figure}

Once a coordinate has been accepted for each sampled mode, the mode contributions are summed and converted into Cartesian coordinates:

\begin{equation}
\mathbf r_{f,i}=\mathbf r_{\mathrm{eq},i}+\sum_{j\in\mathcal{M}}\frac{Q_j\,\mathbf e_{j,i}}{\sqrt{m_i}},
\end{equation}

\noindent where \(\mathbf r_{f,i}\) and \(\mathbf r_{\mathrm{eq},i}\) are the final and equilibrium Cartesian positions of atom \(i\), respectively; \(\mathbf e_{j,i}\) is the eigenvector of mode \(j\) for atom \(i\); \(Q_j\) is the accepted coordinate for mode \(j\); \(m_i\) is the mass of atom \(i\) in atomic mass units, and \(\mathcal{M}\) is the set of positive-frequency non-acoustic modes included in the sampling. The three acoustic modes are excluded. A workflow for the initial conditions generation is shown in Figure \ref{fig:workflow1}.

One of the major drawbacks of this approach is the requirement to generate a substantial number of points to ensure sampling of the entire region of possibilities. In our specific case, this entails generating a large number of new coordinate sets to ensure that all potential displacements are adequately sampled. Consequently, reducing the computational cost is highly desirable. To this end, we use an unsupervised machine-learning pipeline to reduce costs through a clustering-based approach. For this purpose, a cascade strategy was adopted. First, the generated geometries are transformed into a set of atomic descriptors based on the Smooth Overlap of Atomic Positions (SOAP) representation \cite{De2016, Himanen2020}. These descriptors encode the local atomic density around a reference atom using Gaussian functions, from which a unique vector can be constructed for each chemical environment. We then applied kernel principal component analysis (kernel PCA, kPCA), a nonlinear dimensionality reduction technique that maps the original descriptor vectors into a higher-dimensional feature space through a kernel function and extracts the principal components in that space \cite{Scholkopf1998}. In this way, nonlinear relationships in the data can be captured while still providing a lower-dimensional representation of the structures. This dimensionality reduction facilitates dataset handling and enables the grouping of similar structures. The resulting representations are then clustered using the K-means algorithm. More details are provided in the Supportive Information.

The clustering procedure is carried out iteratively. Initially, a set of points is randomly selected as cluster centers. Each data point is then assigned to the nearest center based on the Euclidean distance. The cluster centers are subsequently updated, and the assignment process is repeated until convergence, that is, until the change in cluster membership between successive iterations becomes negligible. The quality of the clustering and the choice of the number of clusters were further assessed using silhouette analysis \cite{Rousseeuw1987}. 

\begin{figure}
    \centering
    \includegraphics[width=0.45\linewidth]{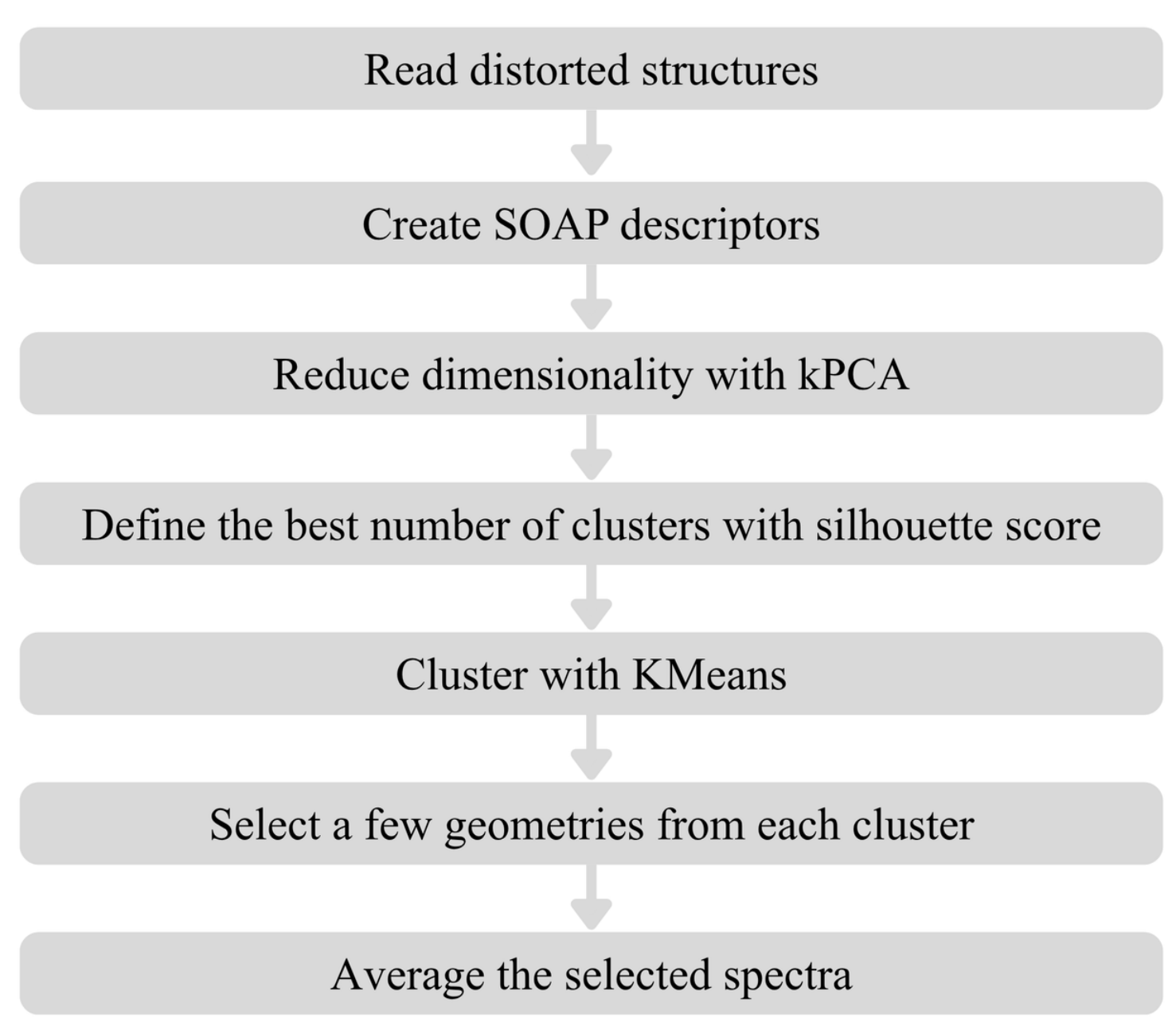}
    \caption{Workflow for the clustering process.}
    \label{fig:workflow2}
\end{figure}

The resulting clusters are used to color the spectra, allowing visual identification of distinct groups. The averaged spectra within each cluster further show that the clustering effectively separates meaningful subsets of the data. A few geometries near each cluster center are selected, and the spectra of all selected geometries are then averaged to obtain the final spectrum. This process is physically motivated by the fact that the optical response is ultimately evaluated as a function of the displaced nuclear geometry. For example, several accepted structures may correspond to different amplitudes along the same normal mode, or to nearly equivalent distortions associated with modes that describe the same type of motion along different directions. In these cases, using all geometries would introduce redundant calculations, while clustering allows us to select a smaller representative subset. A workflow for the clustering process is shown in Figure \ref{fig:workflow2}.

\section{Computational Details}

The VCMC workflow was implemented in Python 3.11 using NumPy 1.26.4, SciPy 1.15.0, ASE 3.26.0, DScribe 2.1.1, scikit-learn 1.6.1, Matplotlib 3.10.0, and tqdm 4.67.1. All electronic-structure calculations were performed using version 6.0 of the Vienna Ab initio Simulation Package (VASP) \cite{Kresse1996a, Kresse1996b, Kresse1999}.

The crystal structures were taken from a CIF file on the Crystallographic Open Database (COD) \cite{Graulis2009}. The atomic positions and unit cell were subsequently optimized. For that purpose, the functional PBE was used with a projected augmented-wave (PAW) contraction. The plane-wave cutoff energy was set to 1.3 times the highest recommended cutoff (ENMAX). Phonons were calculated using density functional perturbation theory (DFPT), with $10^{-8}$ eV as the energy convergence criterion. All the calculated frequencies were real. 

The electronic spectra were calculated from the imaginary part of the frequency-dependent dielectric function via linear response, summing over virtual states \cite{Gajdos2006}. GW-optimized PAWs were adopted, as they provide improved accuracy when virtual states are included. The functional and cutoff energies were kept the same. A k-point mesh grid specific to each system was used: for silicon, a 9x9x9 grid; for GaAs, a 13x13x13 grid; for formaldehyde, a 6x6x11 grid. The number of bands was 3 times that of the ground state, and all calculations were limited to 10 eV. For each system, we generated 1000 geometries by simultaneously sampling all included normal-mode coordinates from their harmonic vibrational distributions at 300 K, with vibrational states selected according to the Boltzmann probabilities, and calculated the spectra for each geometry. The results were processed through the clustering algorithm. The parameters used on this algorith were all tested and results are shown in the supportive information. Principal component analysis was performed using a radial basis function (RBF) kernel. For each system, we selected a fixed budged, and chose a number of geometries proportional to each cluster. They were, then, reaveraged, and replotted. The calculated optical spectra include both Gaussian and Lorentzian broadenings. The Gaussian broadening is controlled by SIGMA, through the Gaussian smearing used in the Brillouin-zone integration, whereas the Lorentzian broadening is controlled by CSHIFT, which introduces a finite imaginary shift in the frequency denominator and moves the poles away from the real axis. For Si and GaAs, we used CSHIFT = 0.01, and for Formaldehyde and Rubrene, we used CSHIFT = 0.001. For all the systems, the SIGMA was set to 0.05. These values were chosen to be as small as possible to ensure high resolution and to ensure that no bands are covered. Unless told otherwise, the spectra were obtained by averaging all the directions of the dielectric tensor. This methodology yields satisfactory results and was chosen as a compromise between accuracy and computational cost \cite{Ahmed2026, Ondracka2017, YasinAhmed2024}. 

In the present implementation, phonons are sampled at the $\Gamma$ point of the conventional cell used for each material. This treatment includes the zone-center high-symmetry vibrational modes and, through the use of a conventional cell, modes that are folded to $\Gamma$ in the primitive-cell description. The optical response is then evaluated by Brillouin-zone integration over the electronic k-points for each distorted geometry. Therefore, the present approach can capture vibronically induced changes in the low-energy optical response within the finite-cell sampling employed here. 

To quantify how accurately the reduced sets reproduce the full nuclear-ensemble spectrum, the deviation of the clustered subset relative to the complete set was quantified using the relative root mean square error,

\begin{equation}
    \mathrm{RRMSE}=100\left[\frac{\sum_{i=1}^{N_e}\left(I_{\mathrm{subset}}(E_i)- I_{\mathrm{all}}(E_i)\right)^2}{\sum_{i=1}^{N_e}I_{\mathrm{all}}(E_i)^2}\right]^{1/2}.
\end{equation}

Here, $I_{\mathrm{subset}}(E_i)$ and $I_{\mathrm{all}}(E_i)$
are the spectral intensities of the reduced-set and full-ensemble
averages, respectively, and $N_e$ is the number of energy-grid points within the selected window. The RRMSE is expressed as a percentage. To estimate the sampling uncertainty of the full-ensemble mean, 1000 bootstrap replicates were generated. Each replicate was constructed by randomly drawing $N_g$ geometries with replacement from the original ensemble of $N_g$ geometries and averaging their spectra. Thus, a geometry could be selected more than once within a replicate, while others could be omitted. Each bootstrap average was compared with the original full ensemble average to obtain the normalized residuals at each energy. Then, the RRMSE of the clustered subset is compared with the uncertainty of the original set.


\section{Results and discussion}


\subsection{Silicon}

\begin{figure}[H]
    \centering
    \includegraphics[width=0.65\linewidth]{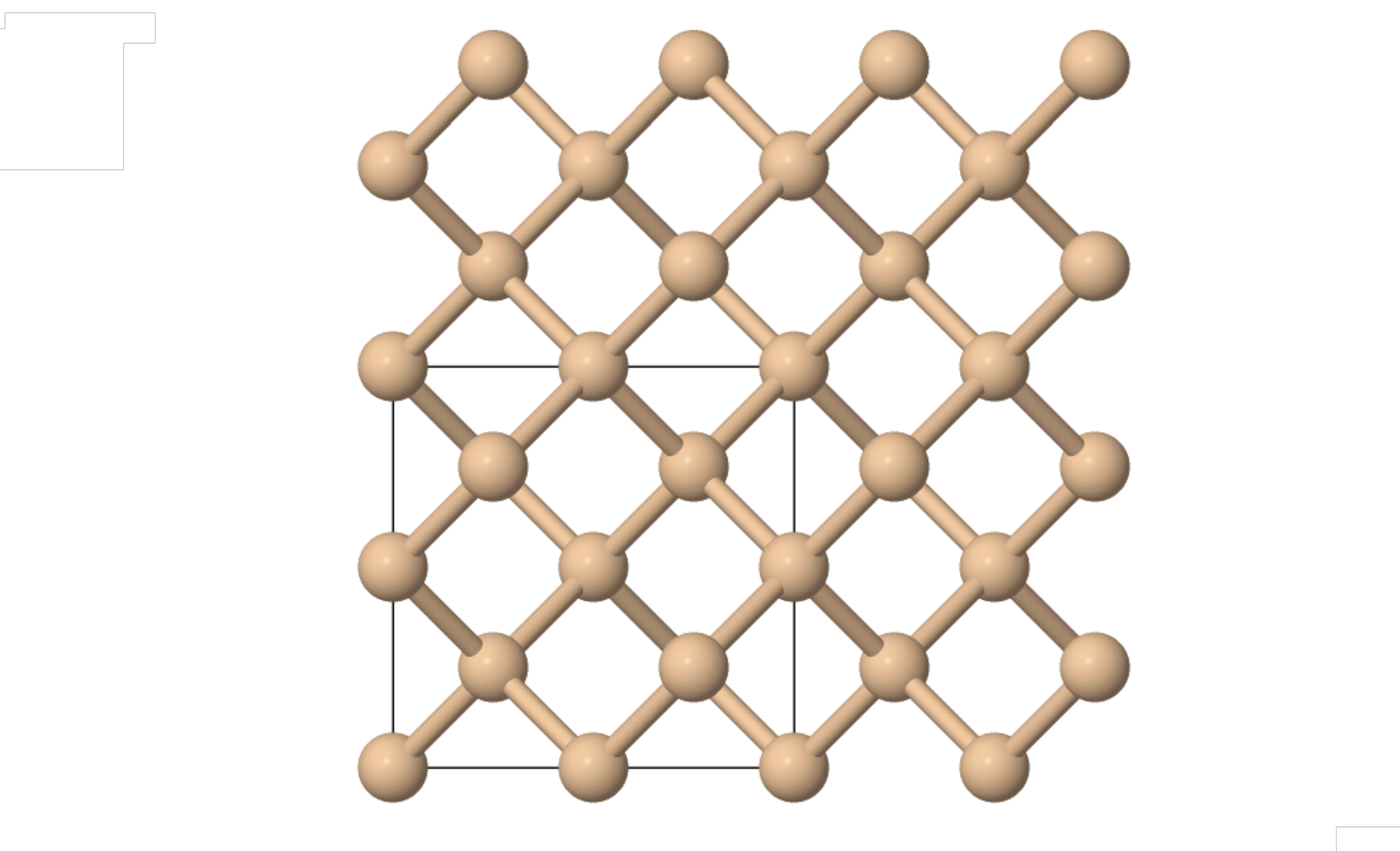}
    \caption{Conventional cubic unit cell of silicon, with the cell boundaries shown in gray. The lattice parameters are $a=b=c=5.47\,\text{\AA}$ and $\alpha=\beta=\gamma=90^\circ$.}
    \label{fig:Si_struct}
\end{figure}

Silicon is a prime example of an indirect-gap semiconductor. Its structure is depicted in Figure \ref{fig:Si_struct}. Experiments show an optical absorption onset around 1.1 eV, corresponding to the indirect band gap, followed by absorption at 3.4 eV, which corresponds to the first direct-gap transition \cite{Noffsinger2012, Zacharias2015}. This indirect band is responsible for the energy harvesting that enables silicon to be widely used in photovoltaics. Therefore, it is very important to simulate it properly. 

\begin{figure}[H]
    \centering
    \includegraphics[width=0.98\linewidth]{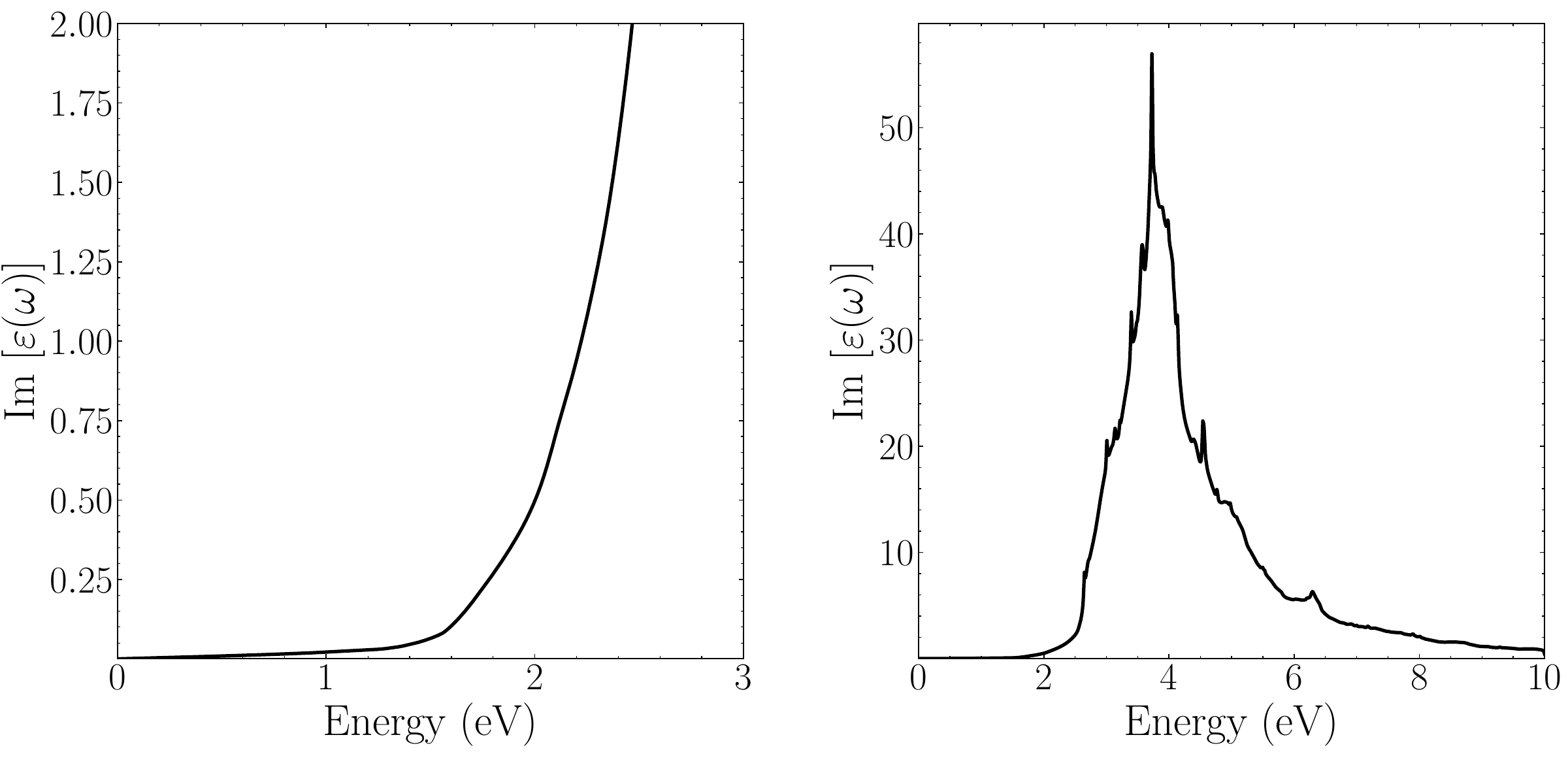}
    \caption{Imaginary part of the dielectric function of silicon, $\varepsilon_2(E)$, calculated at the equilibrium geometry. The left and right panels show the low-energy region and the full plotted energy range, respectively.}
    \label{fig:Si_equil}
\end{figure}

Figure \ref{fig:Si_equil} shows the spectrum calculated at the equilibrium geometry. It displays three main features: a shoulder around 2.6 eV, a stronger band near 3.8 eV, and a weaker high-energy band around 4.5 eV. This overall spectral pattern agrees qualitatively with the experimental dielectric response reported by Aspnes and Studna, which shows corresponding features near 3.5, 4.3, and 5.4 eV \cite{Aspnes1983}. The calculated spectrum, however, is red-shifted relative to the experiment. This deviation is expected, since semilocal functionals such as PBE tend to underestimate band gaps and transition energies in solids, with an error that depends on the material \cite{Borlido2019}. 

\begin{figure}[H]
    \centering
    \includegraphics[width=0.98\linewidth]{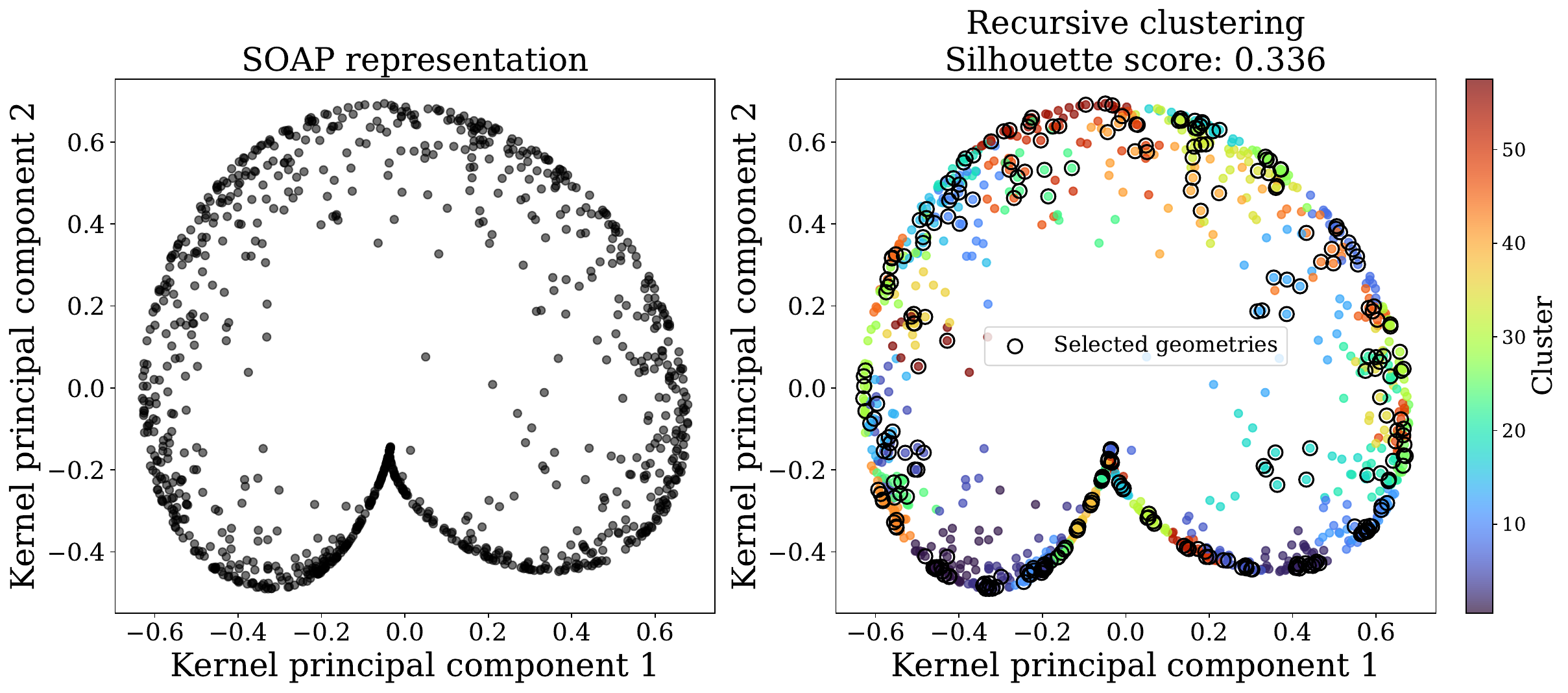}
    \caption{Structural clustering of the sampled silicon geometries in kernel principal component space, showing the distribution of configurations, their cluster assignments, and the selected representative geometries.}
    \label{fig:Si_clustering}
\end{figure}

From the phonon calculation, we obtained 21 real frequencies. As we used the conventional cell, this selection contains the $\Gamma$ points of the primitive cell, but also the $X$ points, which are folded to $\Gamma$ on the simulation cell. The acoustic modes were disregarded because they represent translations and do not change the positions of the atoms relative to one another. The distorted geometries were used to generate new spectra, and all of these geometries were subjected to clustering. The projection of the geometries on principal components, their division into clusters and the clustered selected geometries are shown in the bottom panel of Figure \ref{fig:Si_clustering}. The clustering parameters were chosen following the tests shown in the Supportive Information. 

\begin{figure}[H]
    \centering
    \includegraphics[width=1.0\linewidth]{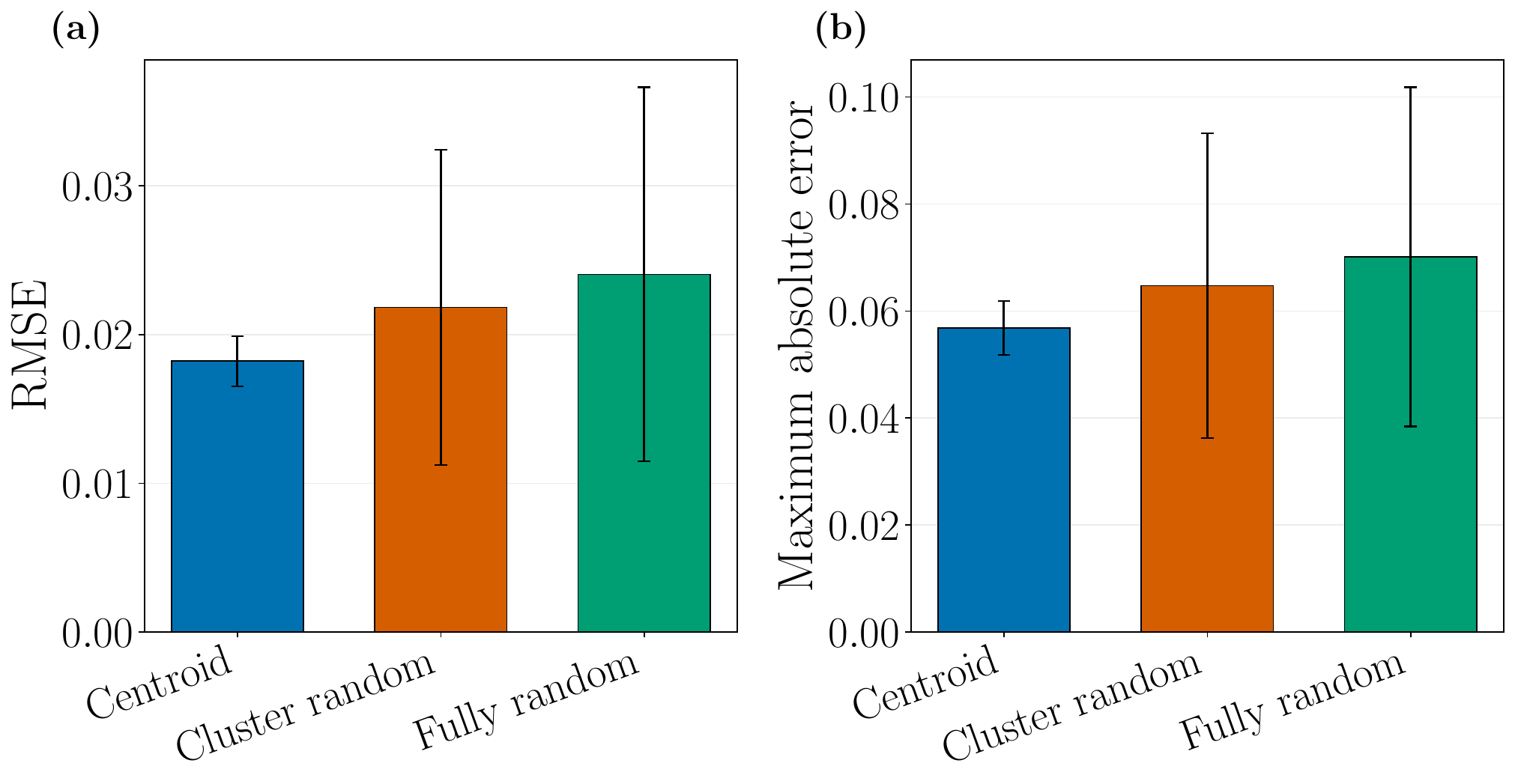}
    \caption{Comparison of representative-structure selection strategies.}
    \label{fig:method}
\end{figure}

Figure \ref{fig:method} compares different ways of sampling the subset. Fully random selection results in a large standard deviation, and a similar behavior is observed when samples are selected randomly within each cluster. However, selecting the samples closest to the cluster centroids considerably reduces the standard deviation, as well as the RMSE and maximum error.

\begin{figure}[H]
    \centering
    \includegraphics[width=0.98\linewidth]{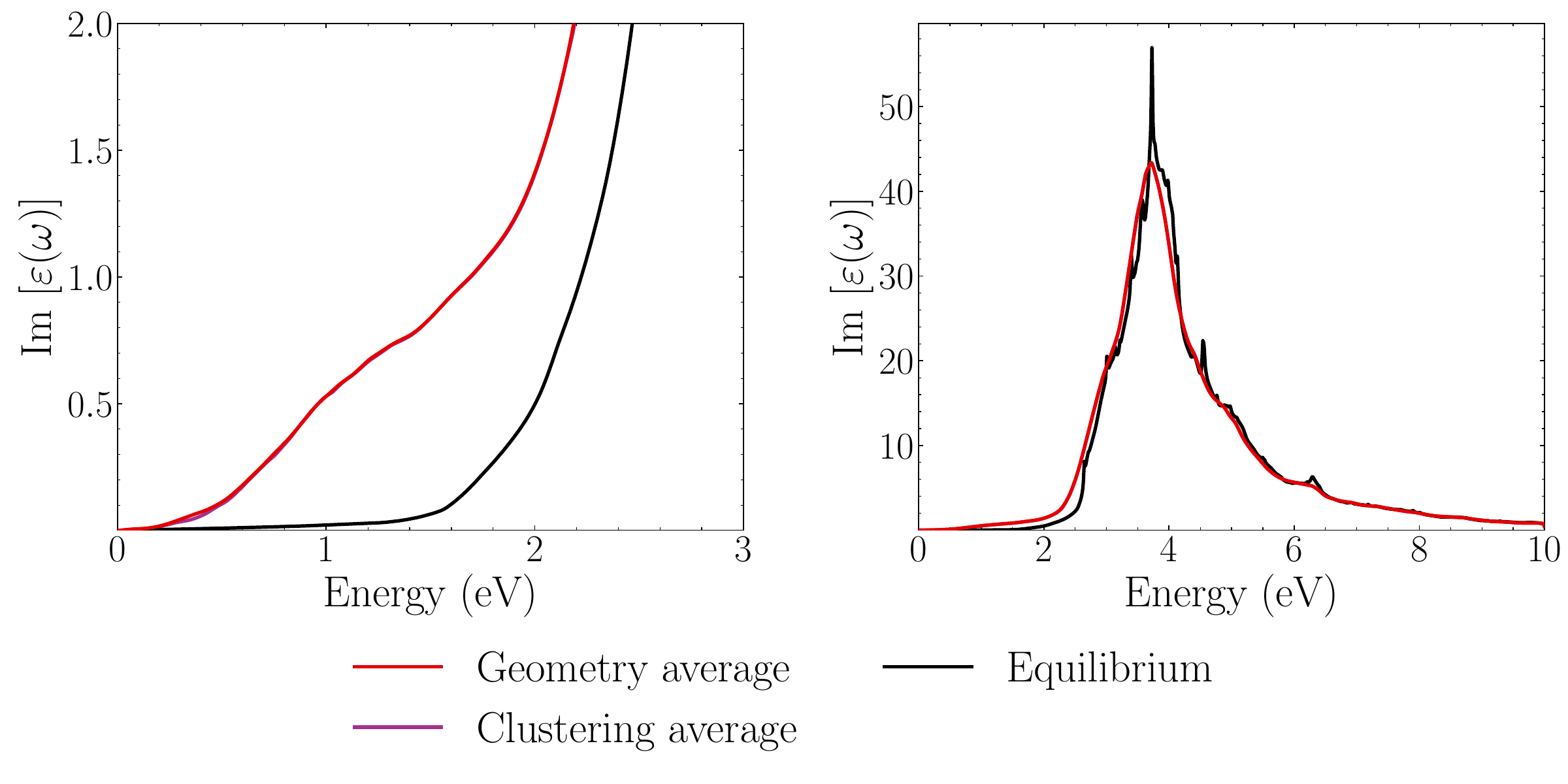}
    \includegraphics[width=1.00\linewidth]{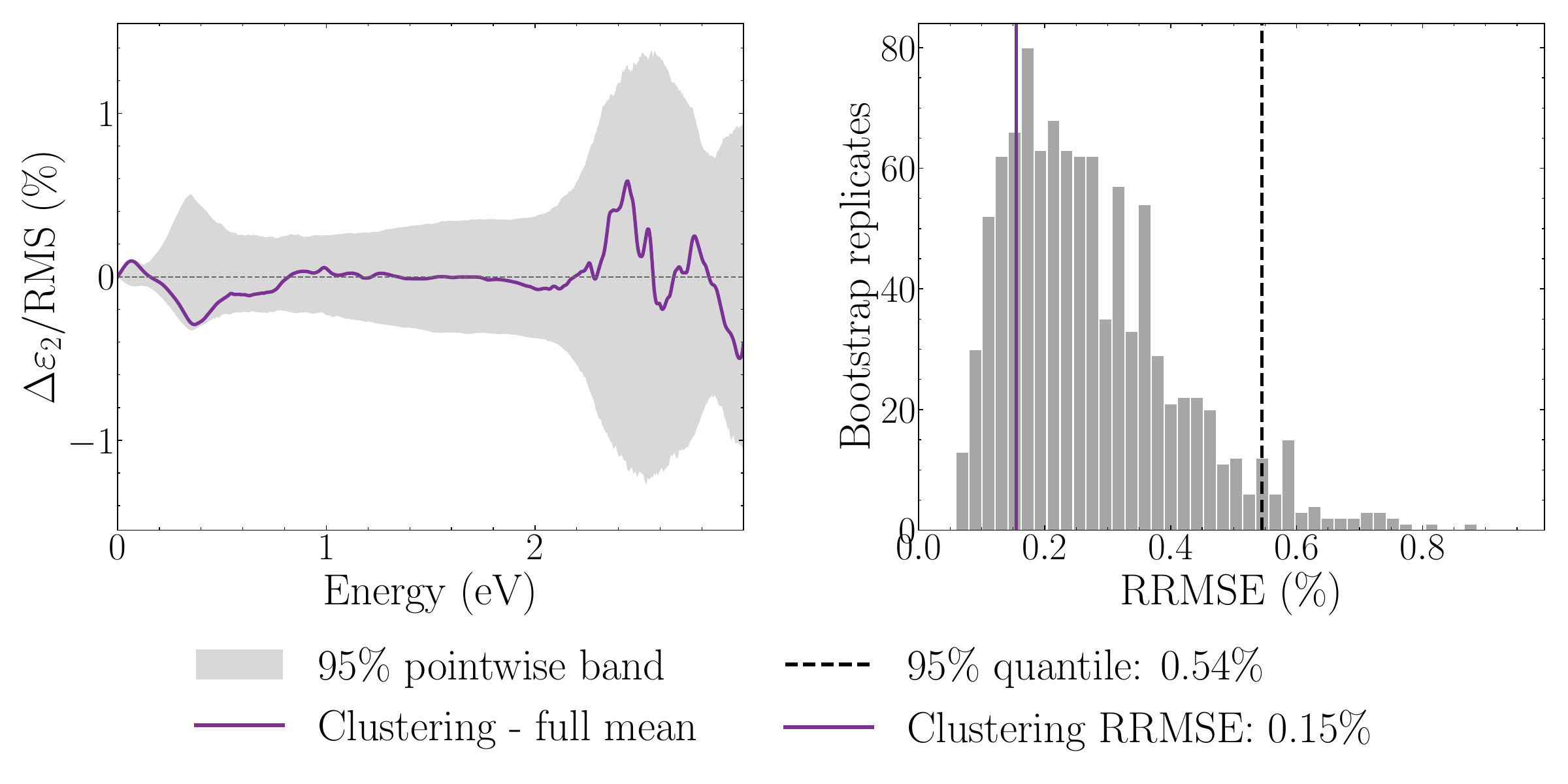}
    \caption{Imaginary part of the dielectric function of silicon, $\varepsilon_2(E)$, and assessment of the cluster-selected subset. The top panels compare the full-ensemble average with the spectrum reconstructed from the cluster-selected geometries and the reference curves identified in the legend. The left and right panels show the low-energy region and the full plotted energy range, respectively. The bottom panels show the signed difference between the cluster-selected and full-ensemble spectra (left) and the corresponding pointwise absolute error (right) over 0-3~eV. Errors were evaluated using unshifted spectra without intensity rescaling.}
    \label{fig:Si_acopvib}
\end{figure}

\begin{figure}[H]
    \centering
    \includegraphics[width=0.98\linewidth]{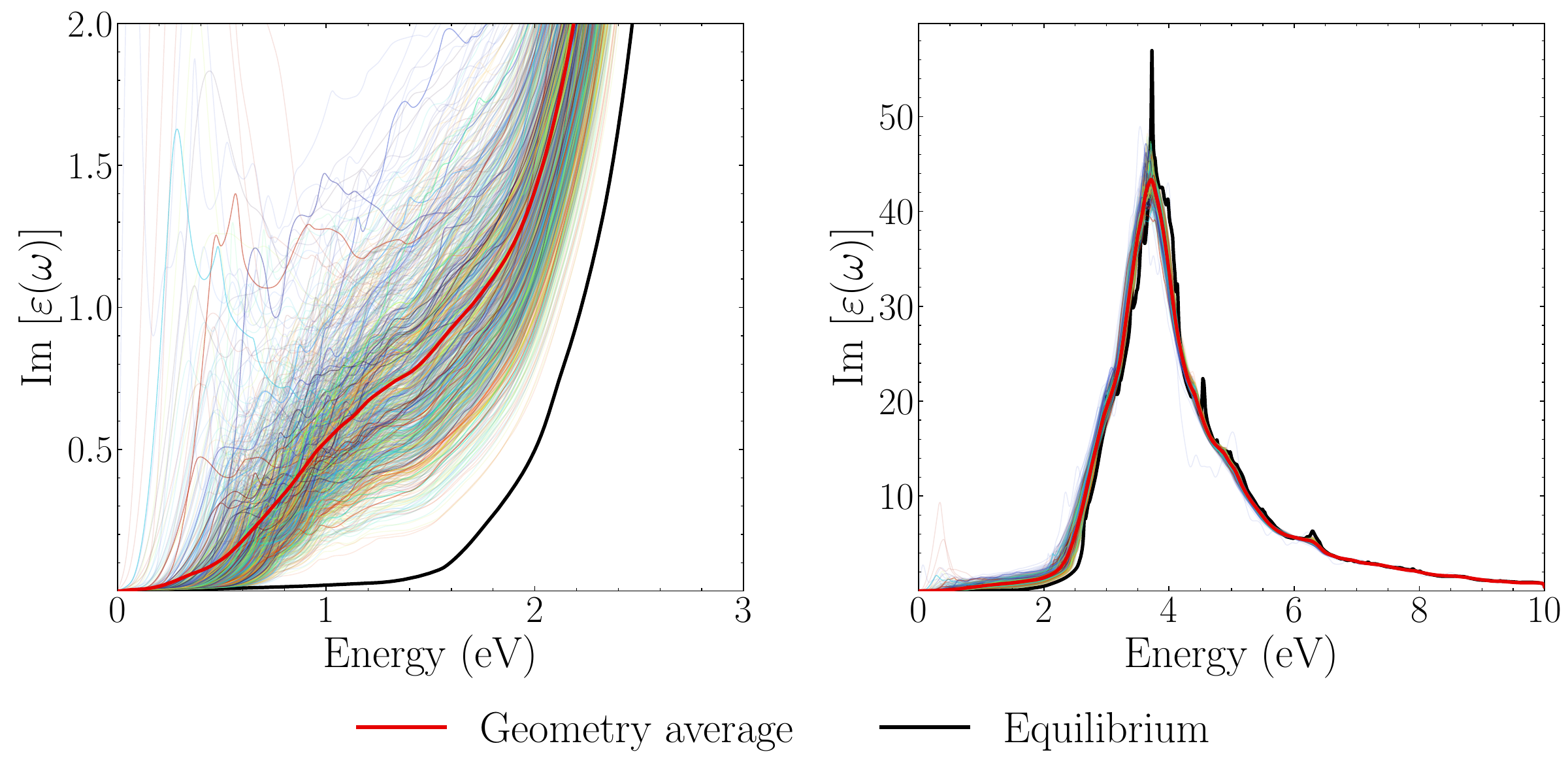}
    \caption{Imaginary part of the dielectric function of silicon, $\varepsilon_2(E)$, calculated for the individual sampled distorted geometries. The left and right panels show the low-energy region and the full plotted energy range, respectively.}
    \label{fig:Si_acopvib_ind}
\end{figure}

The ensemble-averaged spectrum exhibits a low-energy absorption onset near 1 eV, as shown in Figure \ref{fig:Si_acopvib} (top left, red line). The spectrum obtained from the cluster-selected geometries closely follows the average over the full ensemble, as shown in Figure \ref{fig:Si_acopvib} (top left, purple line), indicating that the clustering procedure captures the dominant spectral contributions while using a reduced number of configurations chosen systematically from the distribution of sampled structures, rather than by chance. The individual spectra in Figure \ref{fig:Si_acopvib_ind} show a sizable dispersion among the sampled geometries. 

The agreement between the cluster-selected spectrum and the full-ensemble average was quantified over the 0-3\~eV energy interval using bootstrap resampling. The bottom-left panel of Figure \ref{fig:Si_acopvib} shows the signed spectral difference, normalized by the RMS amplitude of the full-ensemble average, together with the 95\% pointwise bootstrap band. The bottom-right panel shows the bootstrap distribution of the RRMSE. The cluster-selected spectrum yielded an RRMSE of 0.15\%, below the 95th-percentile bootstrap value of 0.54\%. These results indicate that the deviation introduced by the reduced subset is small relative to the estimated sampling uncertainty of the full-ensemble mean.

\begin{figure}[H]
    \centering
    \includegraphics[width=0.98\linewidth]{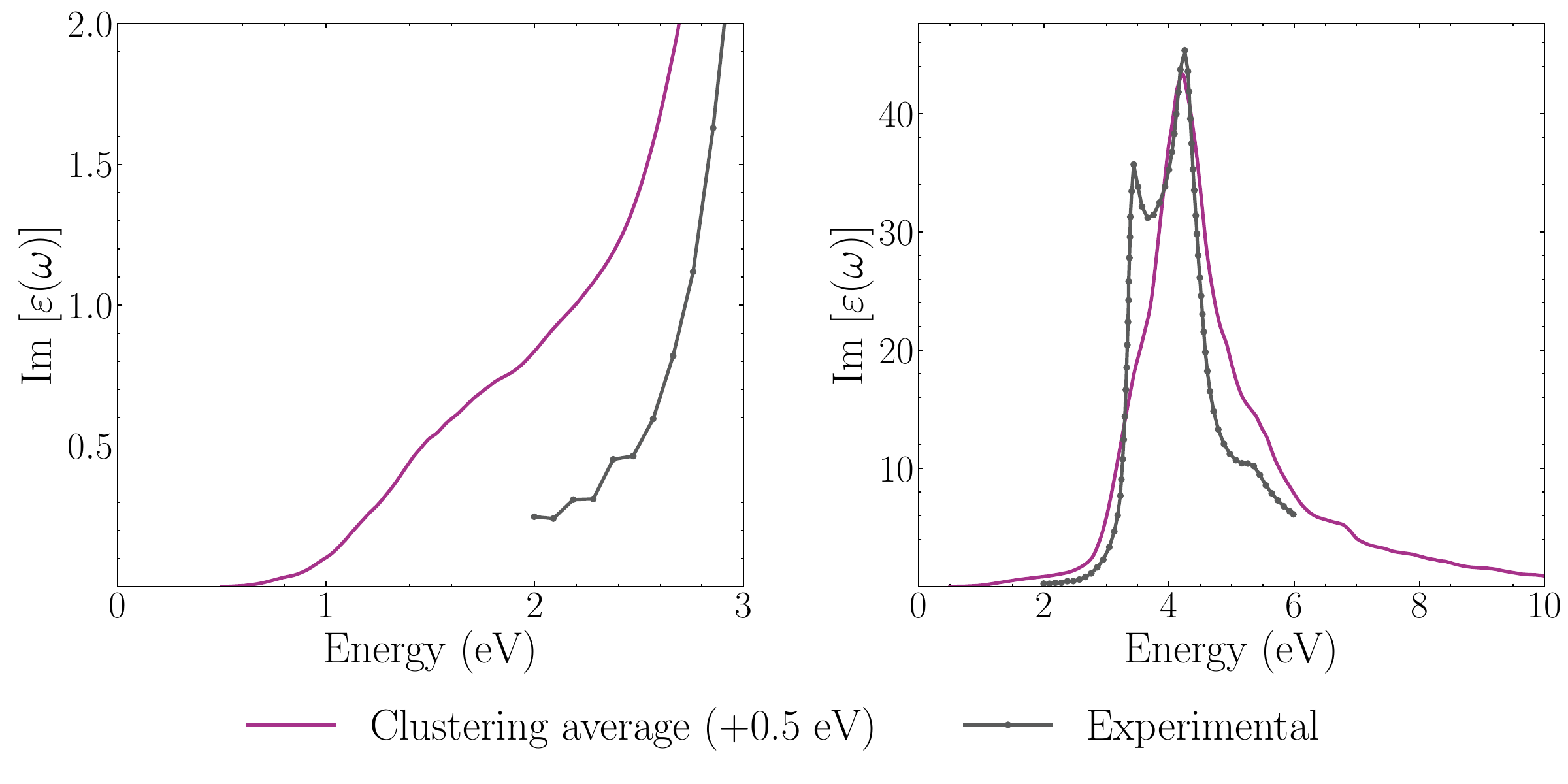}
    \includegraphics[width=0.50\linewidth]{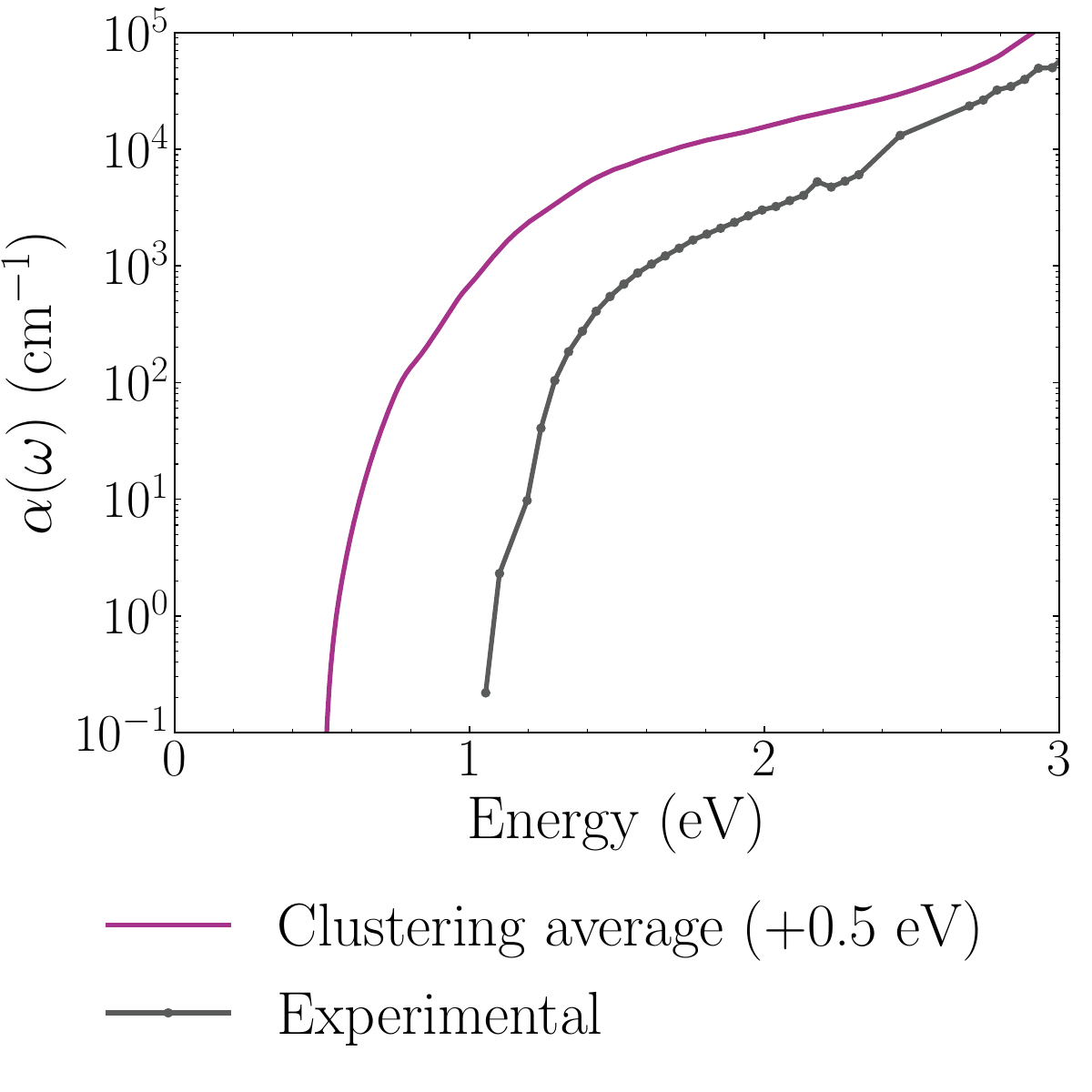}
    \caption{Comparison of the calculated optical response of silicon with experimental data at 300~K. The top panels compare the calculated imaginary part of the dielectric function, $\varepsilon_2(E)$, with the imaginary part of the experimental pseudodielectric function obtained by spectroscopic ellipsometry by Aspnes and Studna \cite{Aspnes1983}. The bottom panel compares the calculated absorption coefficient with the experimental data of Green and Keevers \cite{Green1995}. The calculated spectra were shifted by $+0.50$~eV as an empirical energy alignment.}
    \label{fig:Si_exp}
\end{figure}

For comparison with experiment, the calculated spectra were translated by $+0.50$~eV along the energy axis. This value corresponds to the difference between the equilibrium PBE indirect gap of approximately 0.6~eV and the experimental value of approximately 1.1~eV. The translation is used as an empirical energy alignment for visualization, rather than as a rigorous quasiparticle or scissor correction. Figure~\ref{fig:Si_exp} compares the calculated optical response with experimental data. The experimental dielectric-function data are those reported by Aspnes and Studna \cite{Aspnes1983}, at 300~K, obtained by spectroscopic ellipsometry and expressed as a pseudodielectric function. We also compared the calculated absorption coefficient with the experimental data of Green and Keevers, also at 300 K \cite{Green1995}. This additional comparison, shown in the lower panel of Figure \ref{fig:Si_exp}, shows a good agreement between our calculations and the experimental spectra.

\begin{figure}[H]
    \centering
    \includegraphics[width=0.55\linewidth]{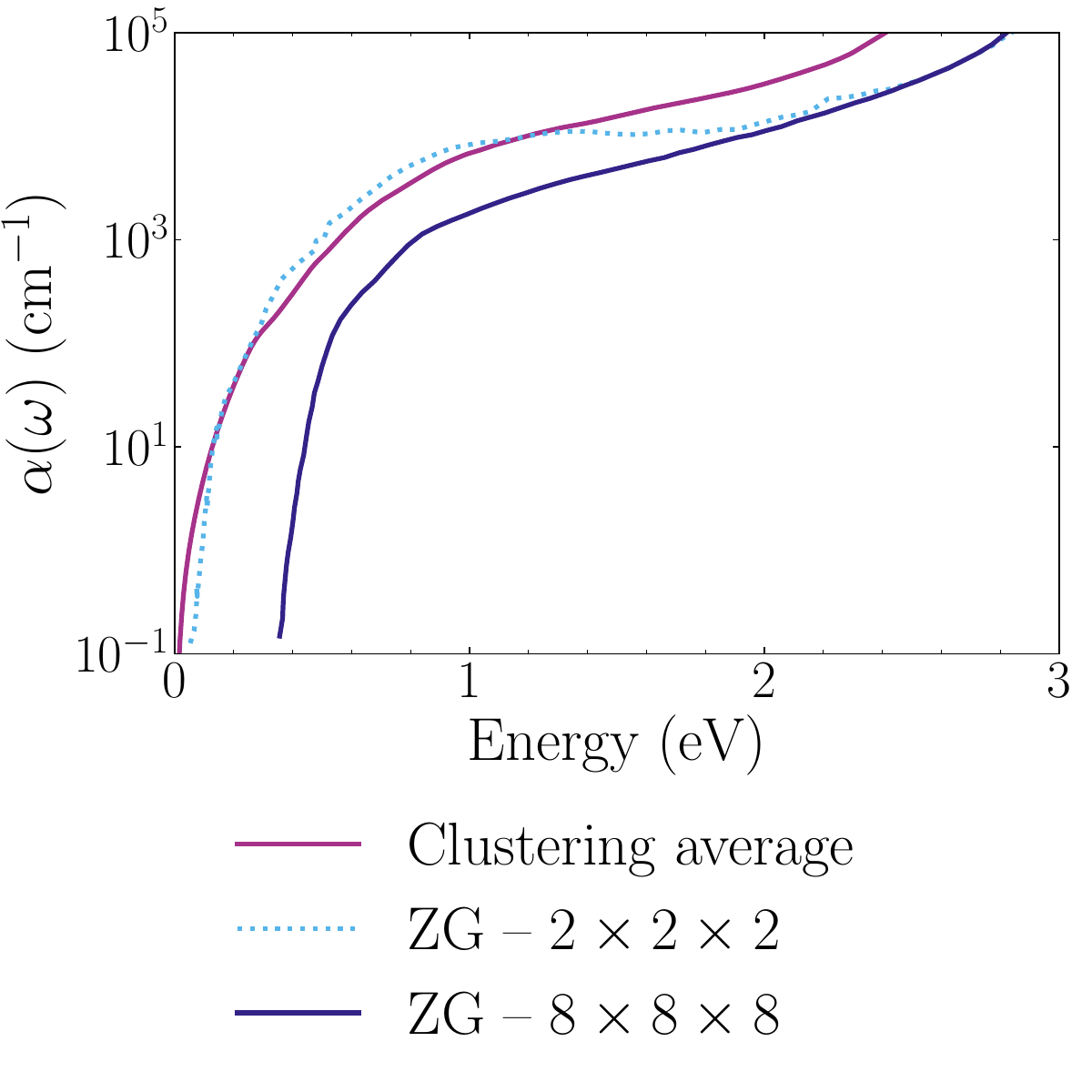}
    \caption{Comparison of the calculated optical response of silicon from the present work with the stochastic Williams-Lax results of Zacharias et al. \cite{Zacharias2015}. The shifts applied by them were removed for proper comparison. The curves are identified in the legend.}
    \label{fig:Si_zg}
\end{figure}

Finally, Figure~\ref{fig:Si_zg} compares the present results with the stochastic Williams–Lax calculations of Zacharias, Patrick, and Giustino \cite{Zacharias2015} at 300~K. In that work, the configurational average was evaluated by importance-sampling Monte Carlo integration. The electronic-structure calculations used LDA within Quantum ESPRESSO. The $2\times2\times2$ and $8\times8\times8$ Born–von Kármán supercell results are reported. These calculations provide a reference for our PBE/VASP results.

\subsection{GaAs}

\begin{figure}[H]
    \centering
    \includegraphics[width=0.65\linewidth]{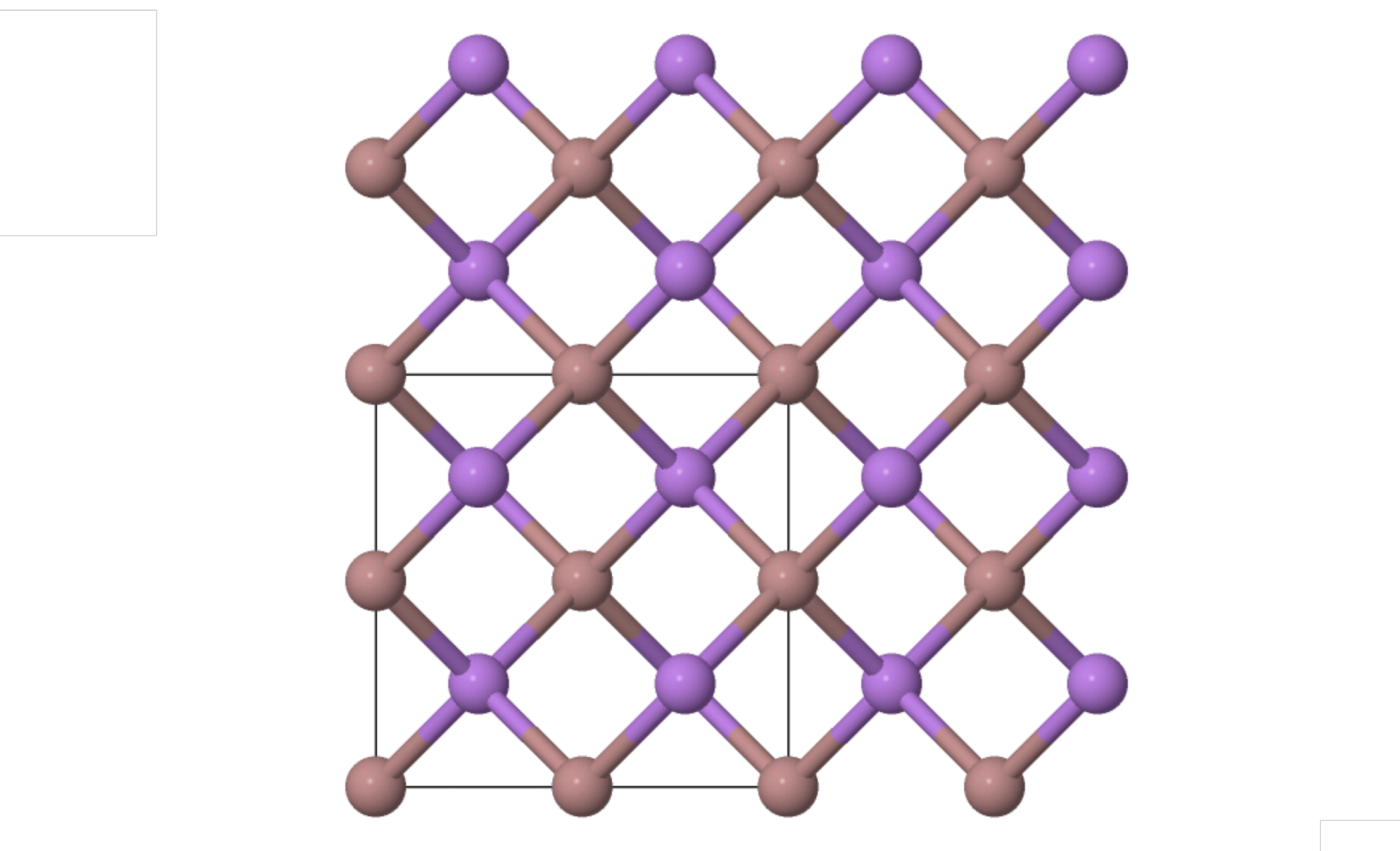}
    \caption{Conventional cubic unit cell of GaAs, with the cell boundaries shown in black. The lattice parameters are $a=b=c=5.76\,\text{\AA}$ and $\alpha=\beta=\gamma=90^\circ$.}
    \label{fig:GaAs_struct}
\end{figure}

The gallium arsenide (GaAs) structure is shown in Figure \ref{fig:GaAs_struct}. Experimentally, it is known to exhibit two bands around 2.9 and 3.1 eV, followed by another band at 4.8 eV. It has a direct band gap of 1.4 eV. This gap does not appear as a distinct band in the spectrum, but it already has a significant intensity \cite{Aspnes1983}. 

\begin{figure}[H]
    \centering
    \includegraphics[width=0.98\linewidth]{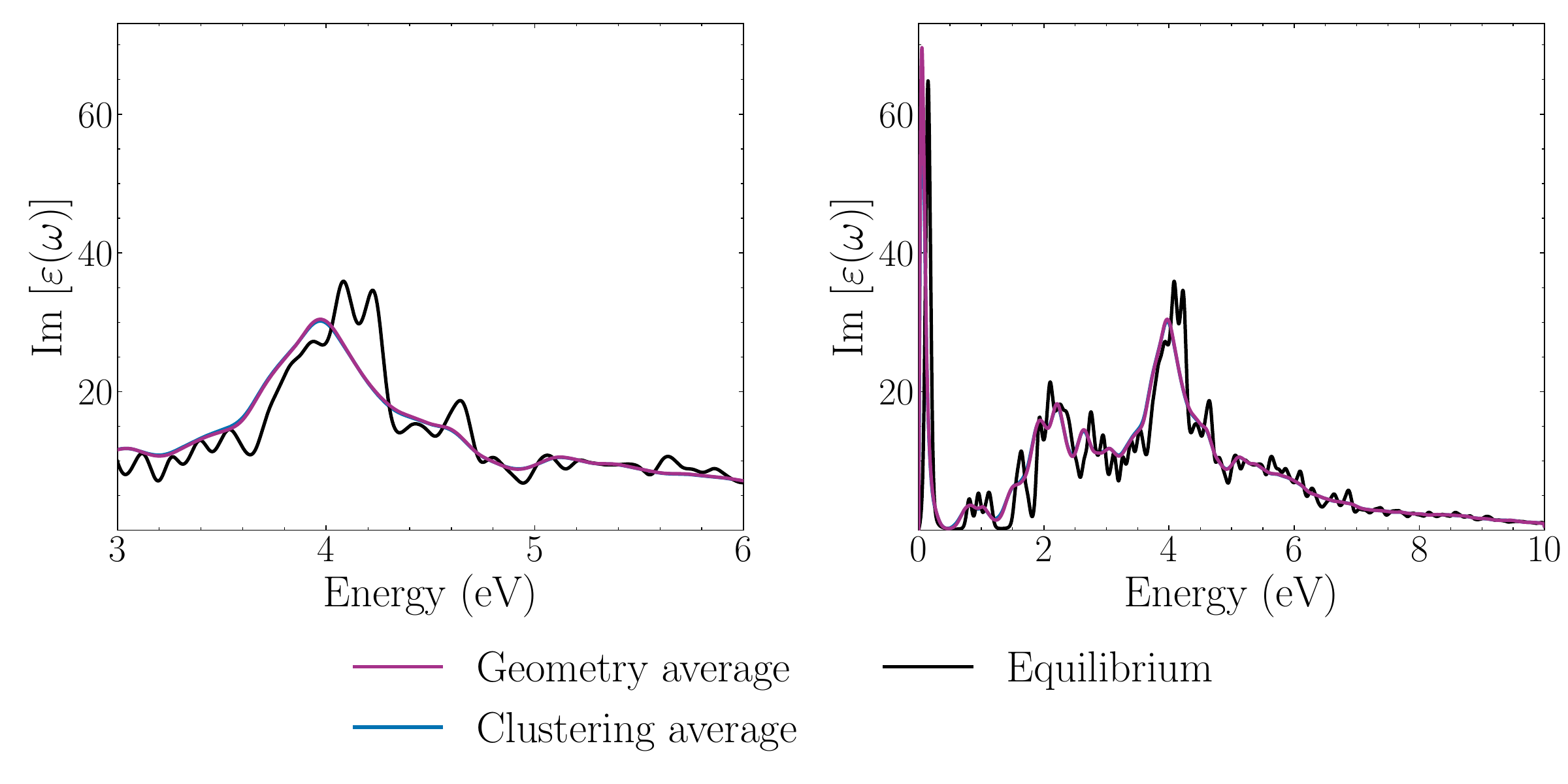}
    \includegraphics[width=1.00\linewidth]{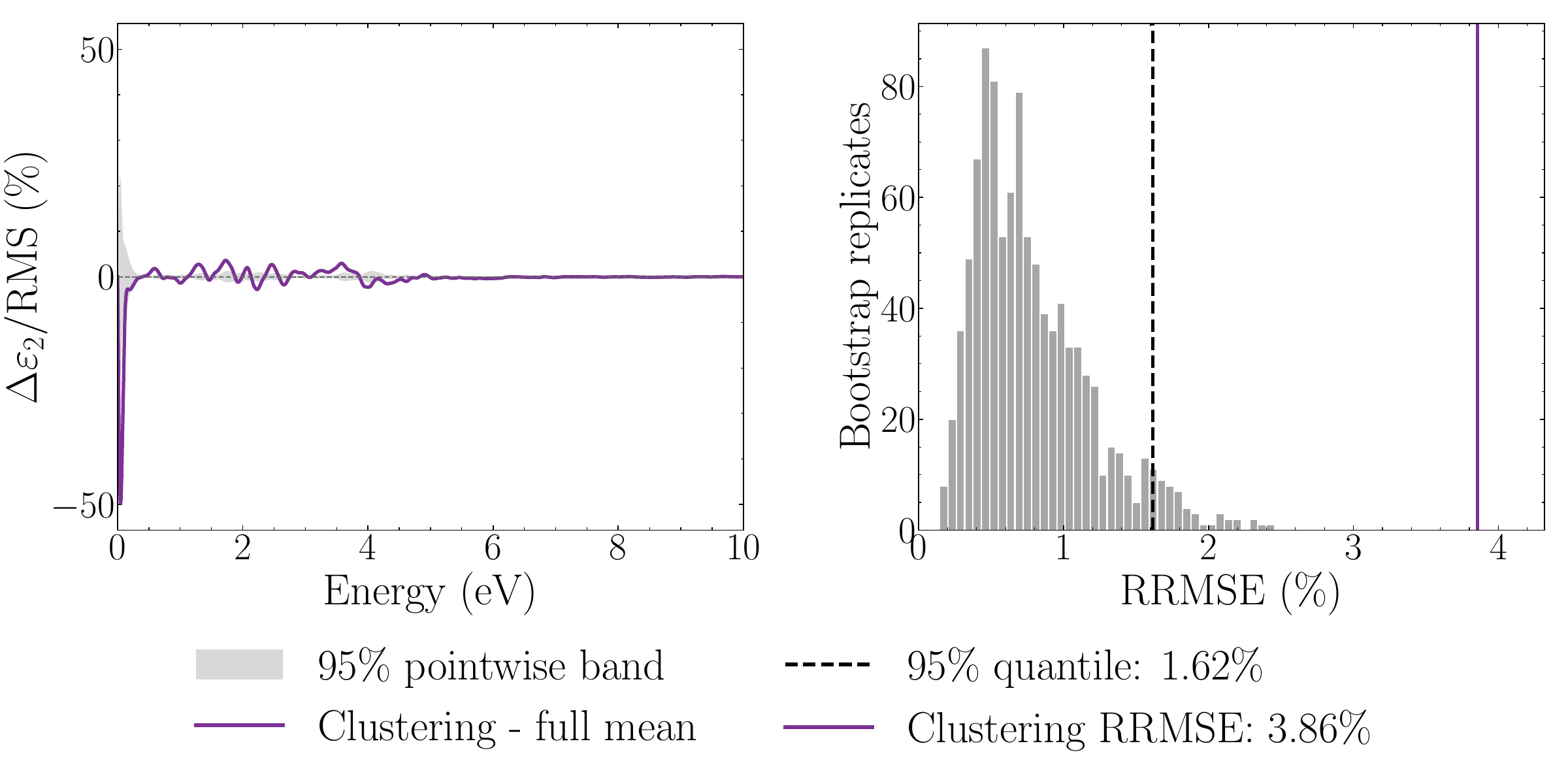}
    \includegraphics[width=0.98\linewidth]{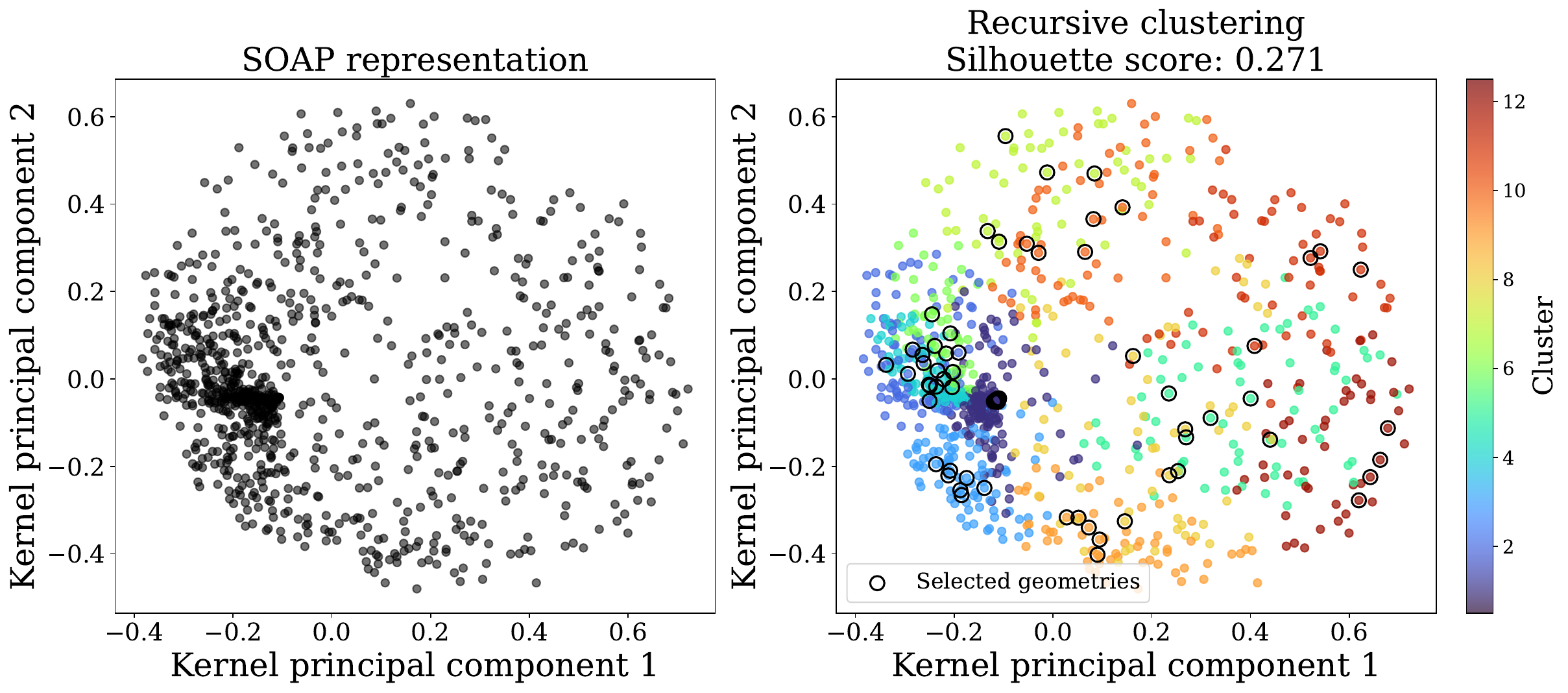}
    \caption{Imaginary part of the dielectric function of GaAs, $\varepsilon_2(E)$, and assessment of the cluster-selected subset. The top panels compare the full-ensemble average with the spectrum reconstructed from the cluster-selected geometries and the reference curves identified in the legend. The left and right panels show the low-energy region and the full plotted energy range, respectively. The middle section presents the error analysis and bootstrap assessment over 0--10~eV. The cluster-selected spectrum has an RRMSE of 3.86\%, compared with a 95th-percentile bootstrap RRMSE of 1.62\%. The bottom section shows the structural clustering in kernel principal component space and the selected representative geometries.}
    \label{fig:GaAs_acopvib}
\end{figure}

The band gap calculated with PBE is 0.15 eV, consistent with the value reported by Wang and co-workers \cite{Wang2026}. Its underestimation relative to experiment reflects the limitations of GGA calculations in describing band gaps \cite{Sham1985, Chan2010, Grossmann2026}, together with the sensitivity of the GaAs band gap to the lattice parameter \cite{Wang2026}. It is, then, expected that the spectra presents a very low energy band with high intensity. The underestimated transition energies contributes to the high intensity of the low-energy band, as $Im[\varepsilon(\omega)]$ has a factor of $1/\omega^2$ \cite{Gajdos2006}, making the low energy bands give a higher contribution to the intensity. Indeed, the equilibrium spectrum, shown in Figure \ref{fig:GaAs_acopvib}, features a very intense band below 0.5 eV, followed by three smaller bands at around 1.0 eV, and increasing bands above 1.5 eV.

\begin{figure}[H]
    \centering
    \includegraphics[width=0.98\linewidth]{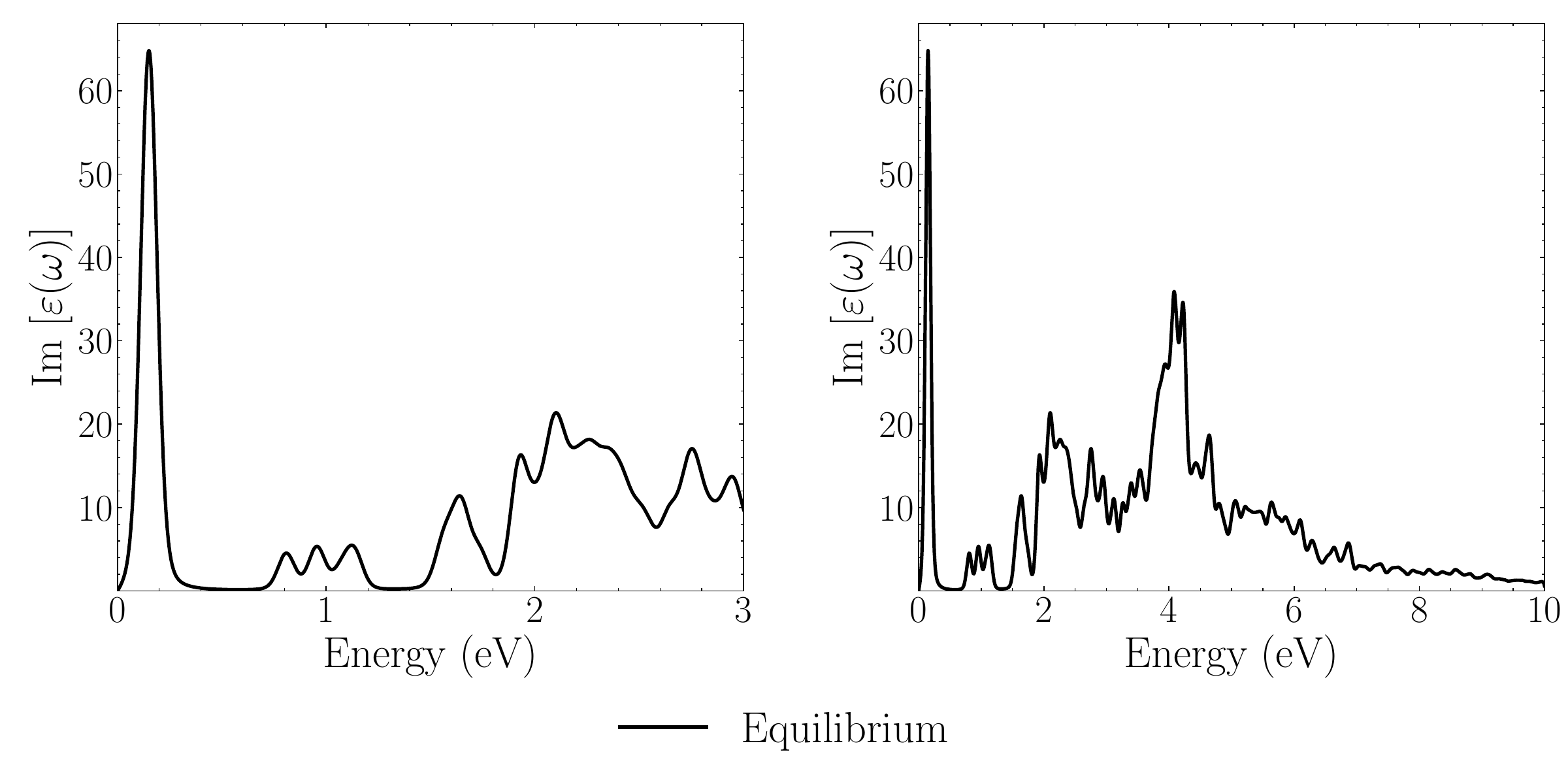}
    \caption{Imaginary part of the dielectric function of GaAs, $\varepsilon_2(E)$, calculated at the equilibrium geometry. The left and right panels show the low-energy region and the full plotted energy range, respectively.}
    \label{fig:GaAs_equil}
\end{figure}

When the spectra of the distorted geometries are averaged, the main effect is a larger broadening of the bands and a change in the relative intensities. It is a effect smaller than in silicon, but it is a very common one in solid systems \cite{Giustino2017, Zacharias2016, Antonius2022, Molina-Snchez2016}. The average of the 1000 points and the clustered average are very similar, with RRMSE of 3.86\%. The 95th percentile of the bootstrap RRMSE distribution was 1.62\%. The error introduced by the reduced subset therefore exceeded this reference by 2.24 percentage points. This discrepancy is seen mainly in the first low energy bands, that shifts a little upon averaging.

\begin{figure}[H]
    \centering
    \includegraphics[width=0.98\linewidth]{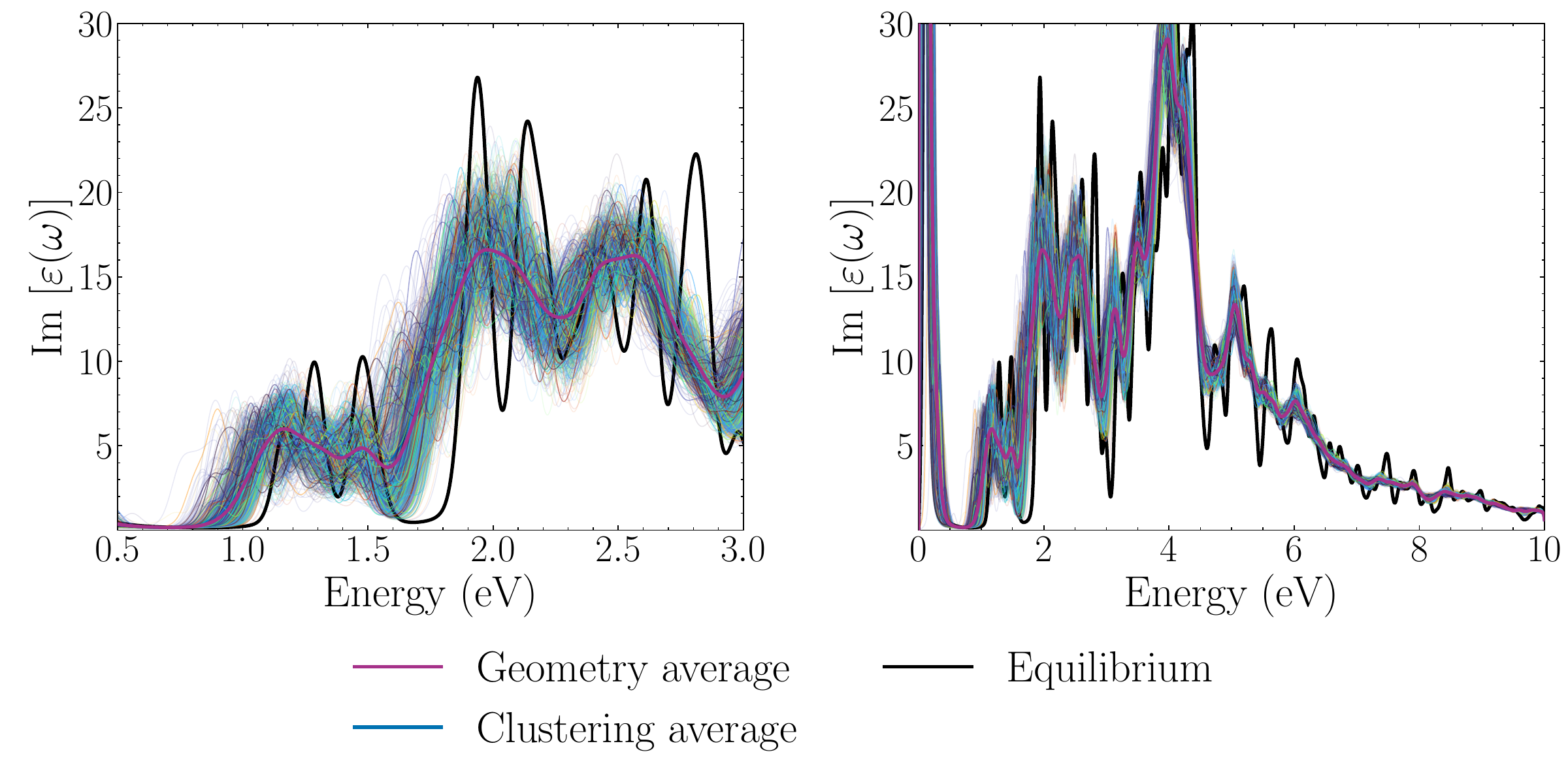}
    \caption{Imaginary part of the dielectric function of GaAs, $\varepsilon_2(E)$, calculated for the individual sampled distorted geometries. The left and right panels show the low-energy region and the full plotted energy range, respectively.}
    \label{fig:GaAg_acopvib_ind}
\end{figure}

When the individual spectra are examined separately, as in Figure \ref{fig:GaAg_acopvib_ind}, the effect of nuclear distortions appears more complex than a simple homogeneous broadening of the band. Different configurations shift spectral intensity both to lower and higher energies relative to the equilibrium spectrum, and these shifts are not fully symmetric. The resulting ensemble average, therefore, reflects an inhomogeneous broadening mechanism, in which the apparent band shape emerges from a distribution of geometry-dependent electronic responses. This analysis shows that the averaged spectrum can hide relevant structures present in the individual configurations.

\begin{figure}[H]
    \centering
    \includegraphics[width=0.98\linewidth]{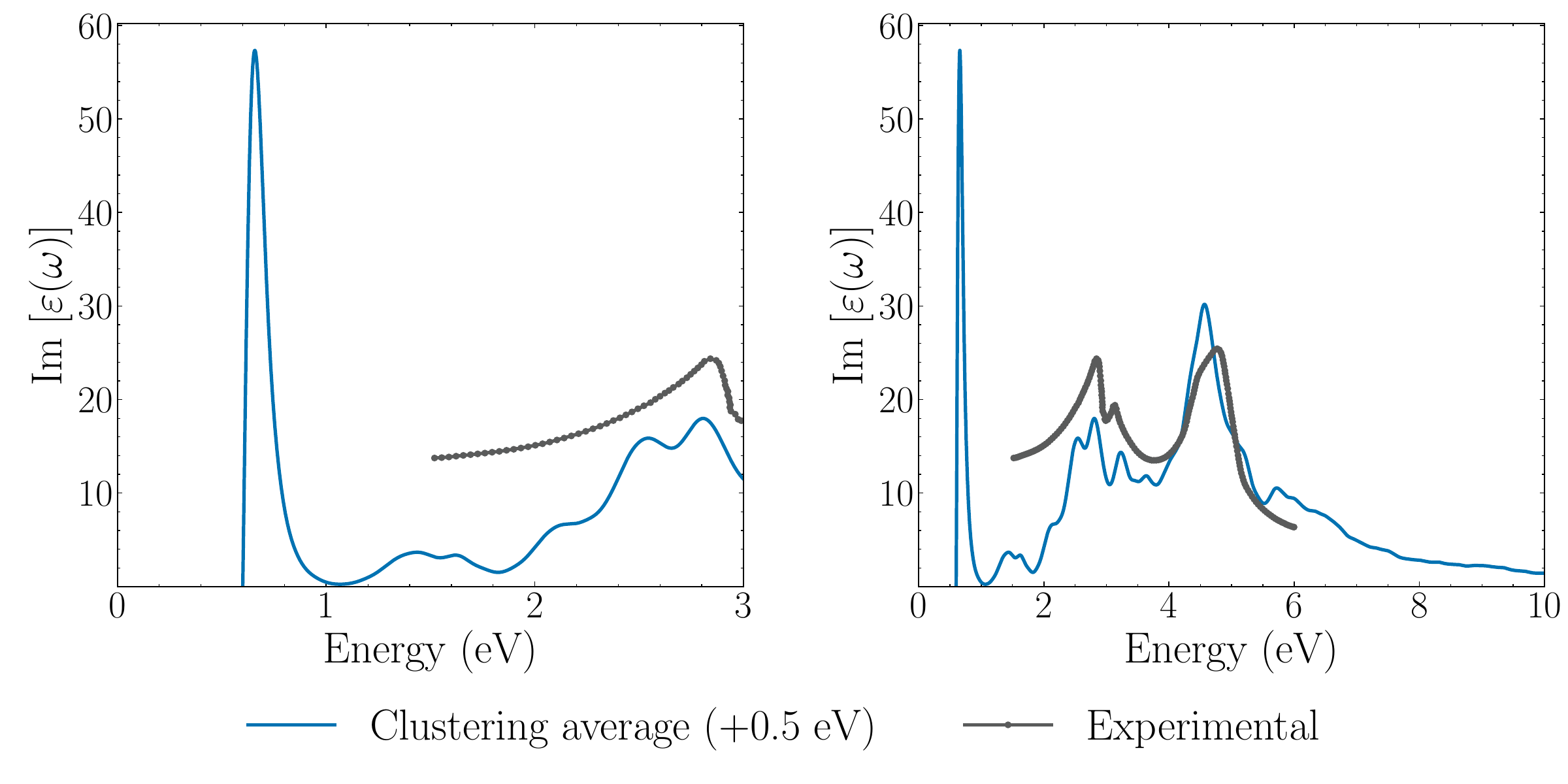}
    \caption{Comparison of the calculated imaginary part of the dielectric function of GaAs, $\varepsilon_2(E)$, with the imaginary part of the experimental pseudodielectric function obtained at 300~K by spectroscopic ellipsometry by Aspnes and Studna \cite{Aspnes1983}. The calculated equilibrium and vibrationally averaged spectra are identified in the legend. The calculated spectra were shifted by $+0.50$~eV as an empirical energy alignment.}
    \label{fig:GaAs_exp}
\end{figure}

For comparison with experiment, the calculated spectra were translated by $+0.50$~eV along the energy axis, used as an empirical energy alignment for visualization. Figure~\ref{fig:GaAs_exp} compares the calculated optical response with experimental data. The experimental dielectric-function data are those reported by Aspnes and Studna \cite{Aspnes1983}, at 300~K, obtained by spectroscopic ellipsometry and expressed as a pseudodielectric function. The calculation captures the main experimental features. It also displays additional structure between the main peaks, suggesting contributions that are not individually resolved in the experimental spectrum.

\subsection{Formaldehyde}

Formaldehyde is a well-known case of vibronic coupling in the gas phase. The isolated molecule has a planar geometry with \ce{C_{2v}} symmetry. Also, it presents a dipole forbidden \ce{n \rightarrow \pi^{*}} electronic transition, characterized as singlet \( A_1 \rightarrow A_2 \) around 4 eV, which has already been broadly studied \cite{Borges2005b, Torres2022, Rocha2000, Roche1974, Dallos2001, Gratien2007}. In the gas phase, it has 6 vibrational modes, with three of them responsible for inducing the vibronic coupling mechanism: two in-plane C-H antisymmetric stretching and bending, and one out-of-plane bending \cite{Borges2005a}. The same vibrational modes are observed in our calculations.

\begin{figure}[H]
    \centering
    \includegraphics[width=0.65\linewidth]{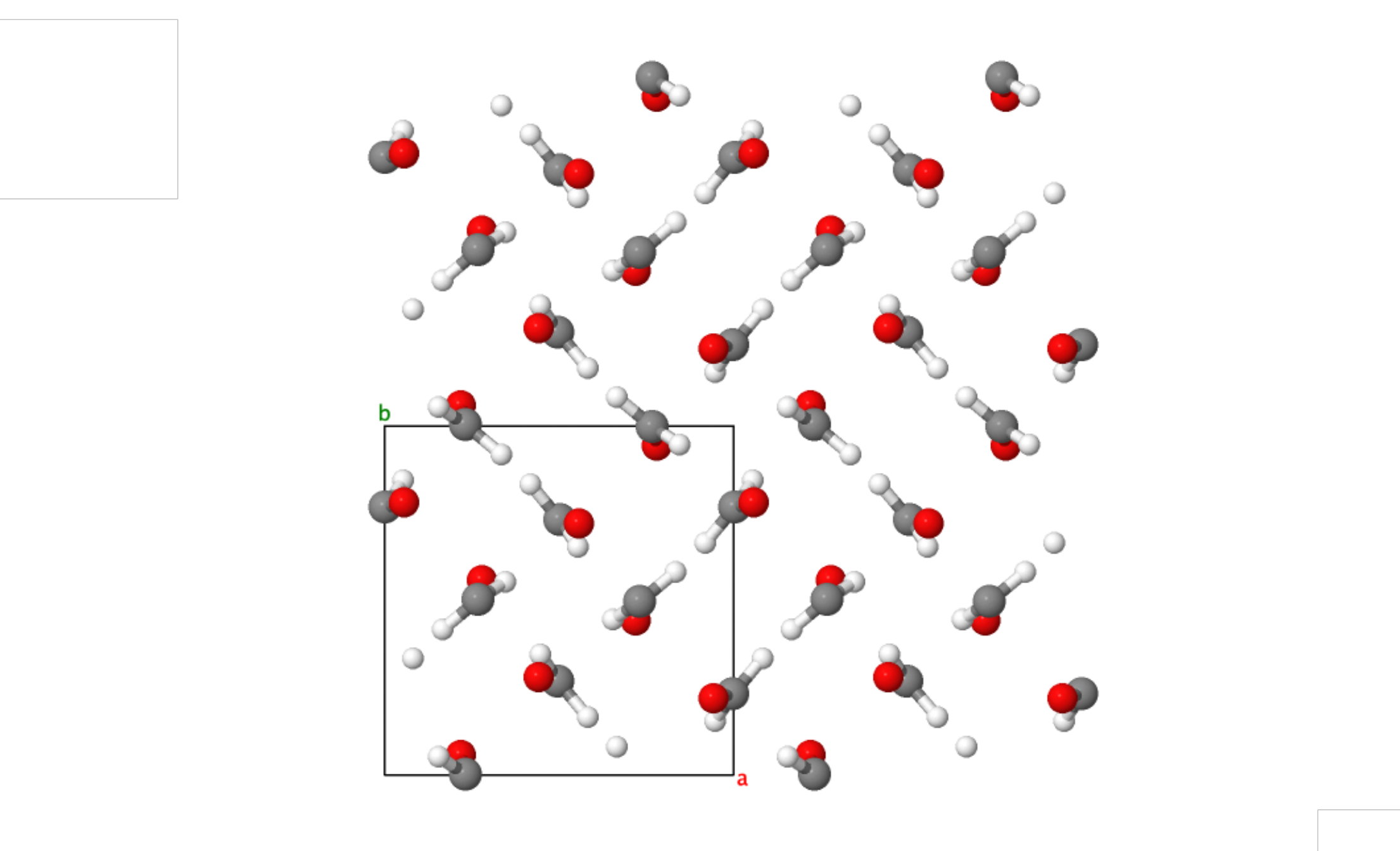}
    \caption{Tetragonal unit cell of crystalline formaldehyde, with the cell boundaries shown in black. The lattice parameters are $a=b=8.03\,\text{\AA}$, $c=4.48\,\text{\AA}$, and $\alpha=\beta=\gamma=90^\circ$.}
    \label{fig:Form_struct}
\end{figure}

In the crystalline form, as shown in Figure \ref{fig:Form_struct}, the molecular units are held by weak \(\ce{C-H}\cdots\ce{O}\) interactions, with interaction energies on the order of 3 kcal/mol \cite{Thakur2011}, causing only minor perturbations on the molecular geometry and electronic structure. Besides, when a crystal contains more than one of the same molecular unit in the cell, it suffers a Davydov splitting. However, the extent of splitting is dependent on the transition dipole moment of each unit \cite{Davydov1971}. Because formaldehyde has a dipole-forbidden transition, the splitting in this case is negligible. As a result, although the crystal environment lowers the molecular symmetry, the vibrational frequencies and electronic transitions of the isolated molecule remain a good approximation for those of the crystal, allowing the crystalline spectrum to be interpreted in terms of the molecular ones. 

\begin{figure}[H]
    \centering
    \includegraphics[width=0.98\linewidth]{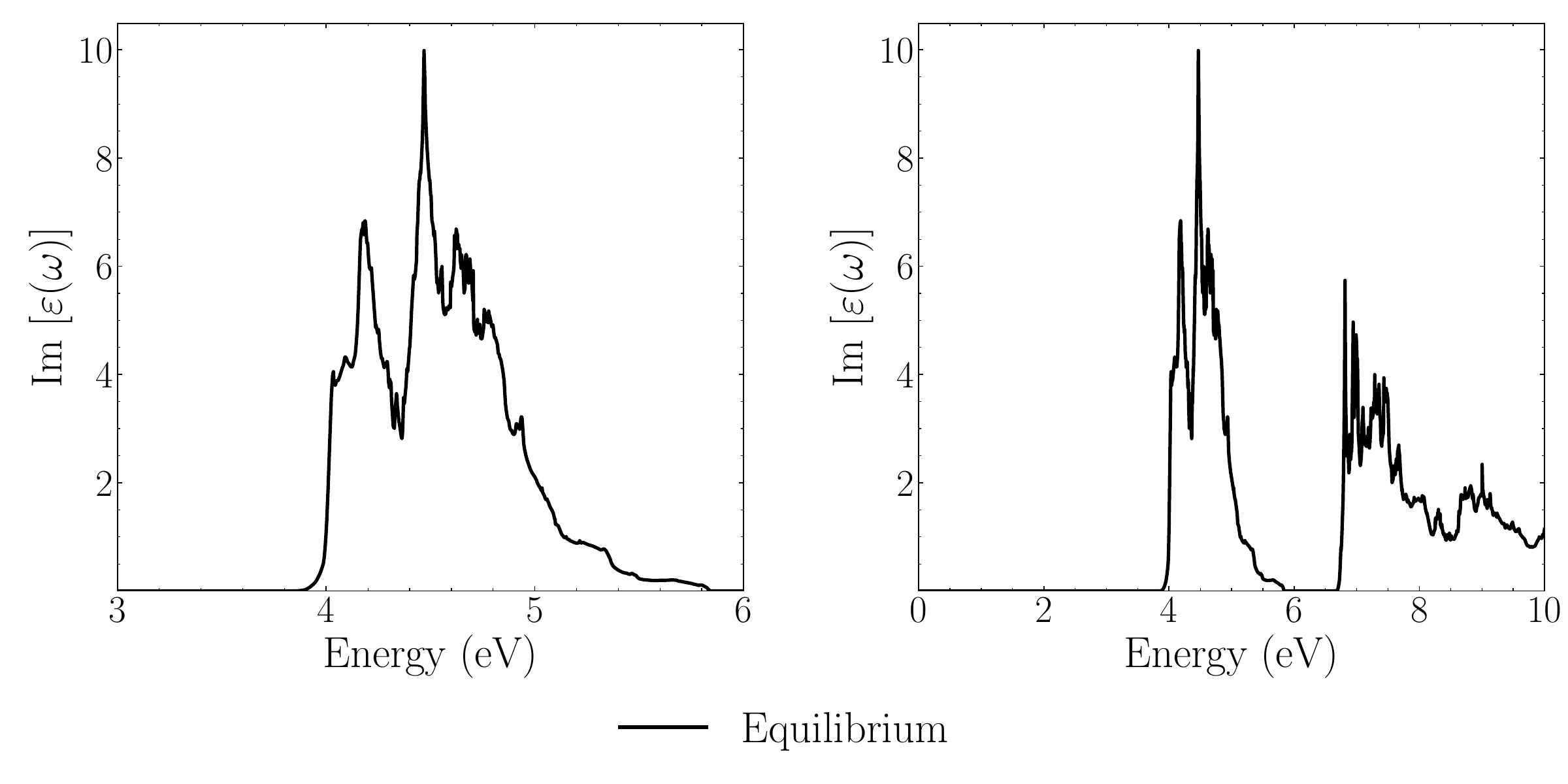}
    \caption{Imaginary part of the dielectric function of crystalline formaldehyde, $\varepsilon_2(E)$, calculated at the equilibrium geometry. The left and right panels show the low-energy region and the full plotted energy range, respectively.}
    \label{fig:Form_equil}
\end{figure}

Figure \ref{fig:Form_equil} shows that the equilibrium geometry exhibits a band that begins at 4 eV, and other bands that begin at almost 7 eV. 

\begin{figure}[H]
    \centering
    \includegraphics[width=0.98\linewidth]{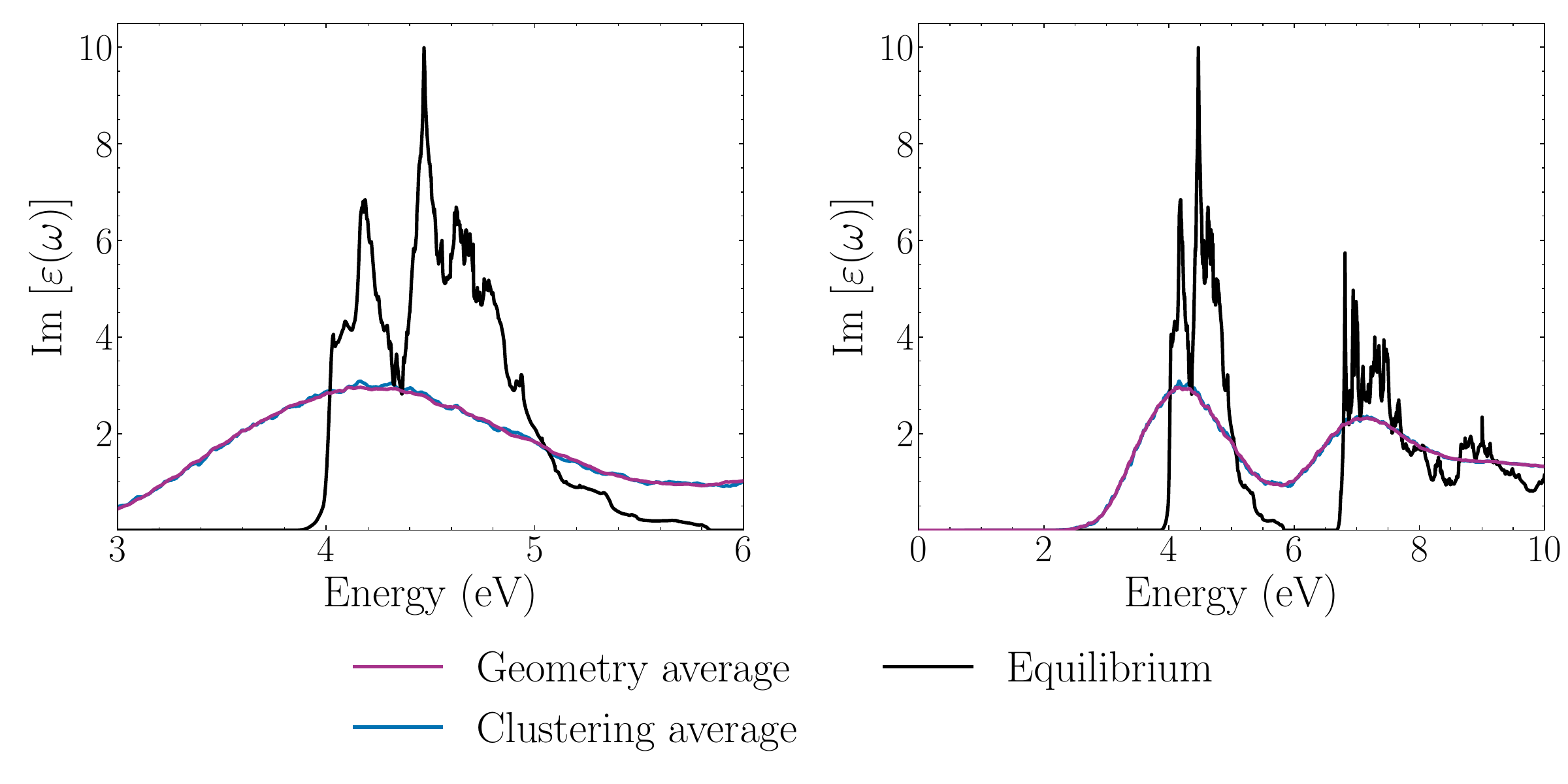}
    \includegraphics[width=0.98\linewidth]{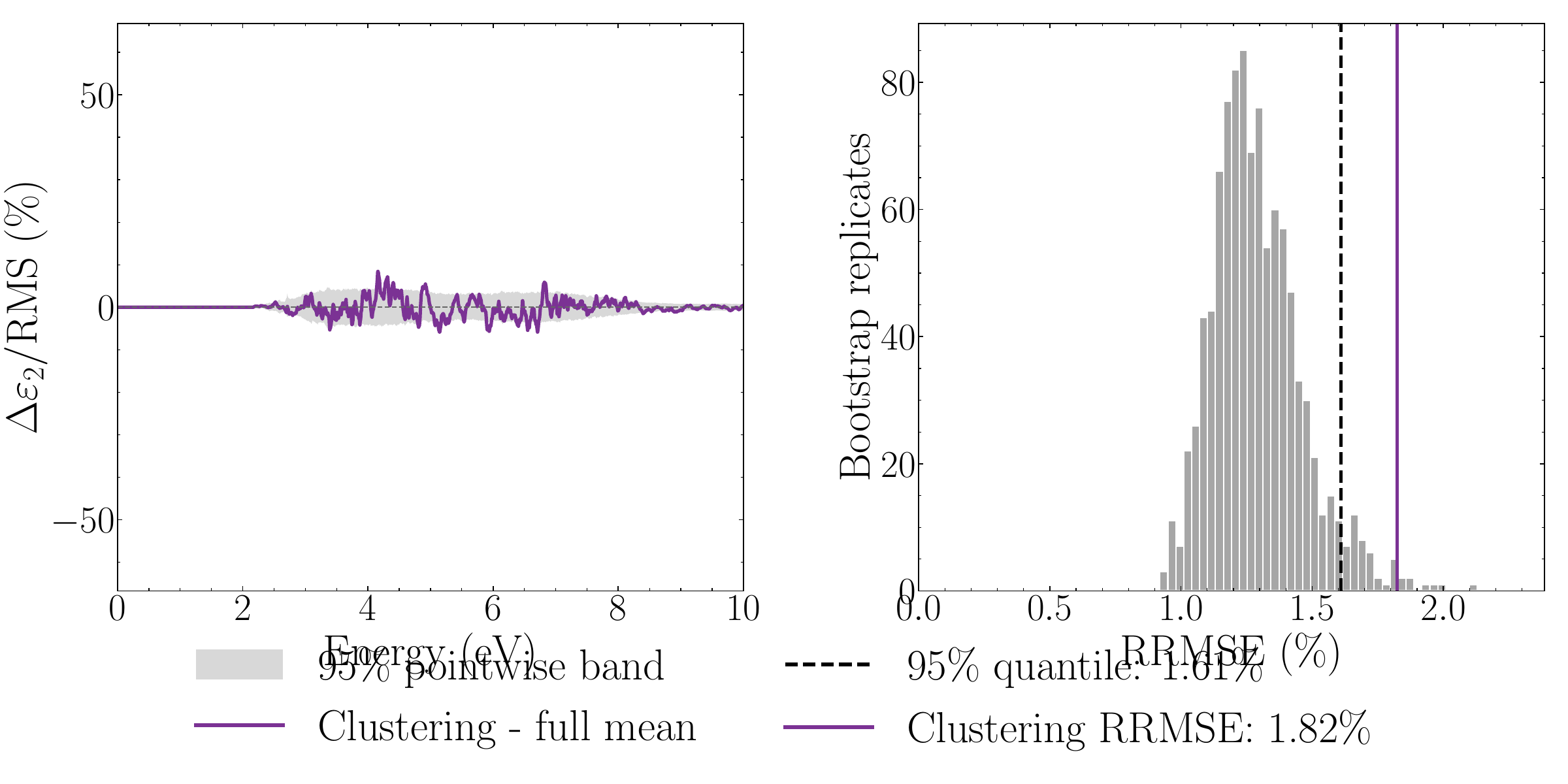}
    \includegraphics[width=0.98\linewidth]{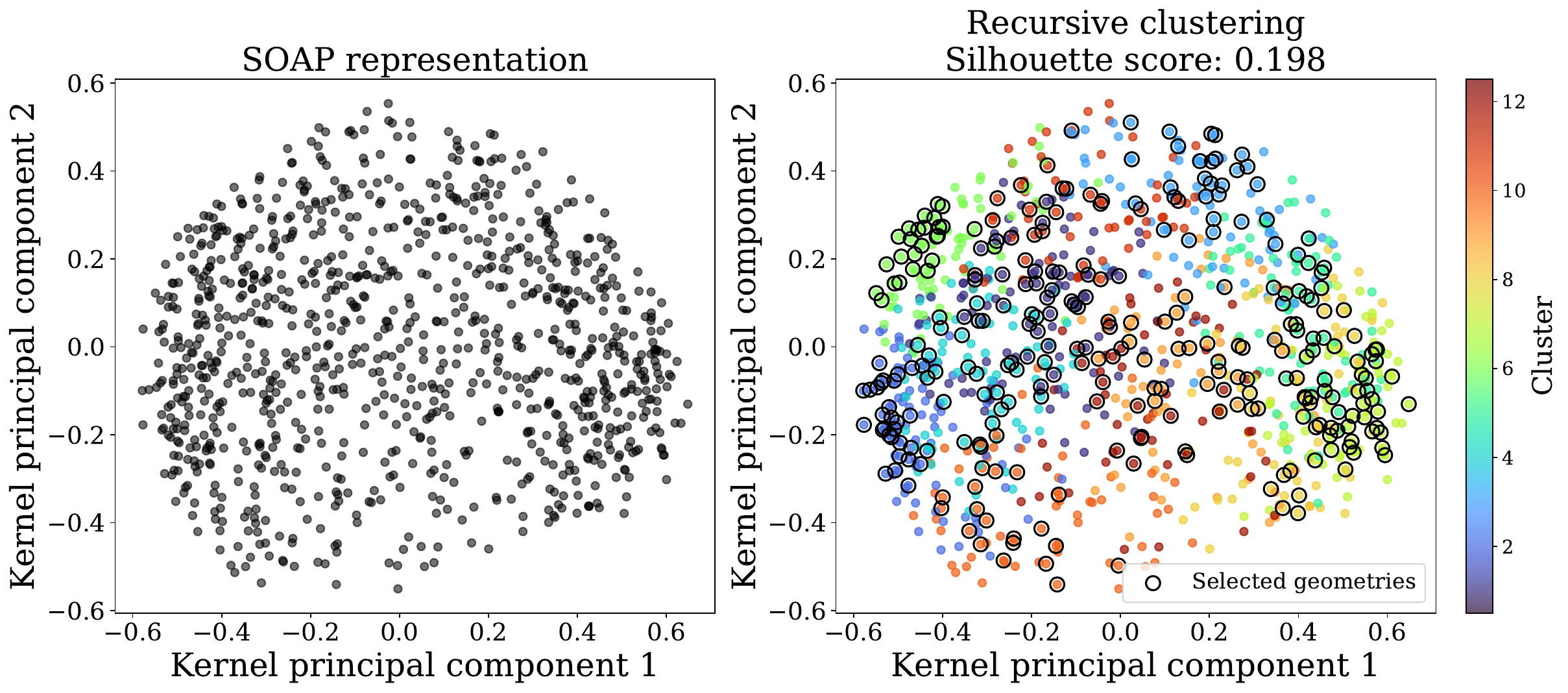}
    \caption{Imaginary part of the dielectric function of crystalline formaldehyde, $\varepsilon_2(E)$, and assessment of the cluster-selected subset. The top panels compare the full-ensemble average with the spectrum reconstructed from 300 cluster-selected geometries and the reference curves identified in the legend. The left and right panels show the low-energy region and the full plotted energy range, respectively. The middle section presents the error analysis and bootstrap assessment over 0--10~eV. The cluster-selected spectrum has an RRMSE of 1.82\%, compared with a 95th-percentile bootstrap RRMSE of 1.61\%. The bottom section shows the structural clustering in kernel principal component space and the selected representative geometries.}
    \label{fig:Form_acopvib}
\end{figure}

Using a budget of 300 geometries, the cluster-selected subset reproduced the full-ensemble average spectrum with an RRMSE of 1.82\%, as shown in Figure \ref{fig:Form_acopvib}. The 95th percentile of the bootstrap RRMSE distribution was 1.61\%, providing a reference for the sampling uncertainty of the full-ensemble mean. The error introduced by the reduced subset therefore exceeded this reference by only 0.21 percentage points, and a reduction of 70\% in the number of geometries, indicating that the deviation associated with geometry selection was comparable in magnitude to the estimated sampling uncertainty. 

\begin{figure}[H]
    \centering
    \includegraphics[width=0.98\linewidth]{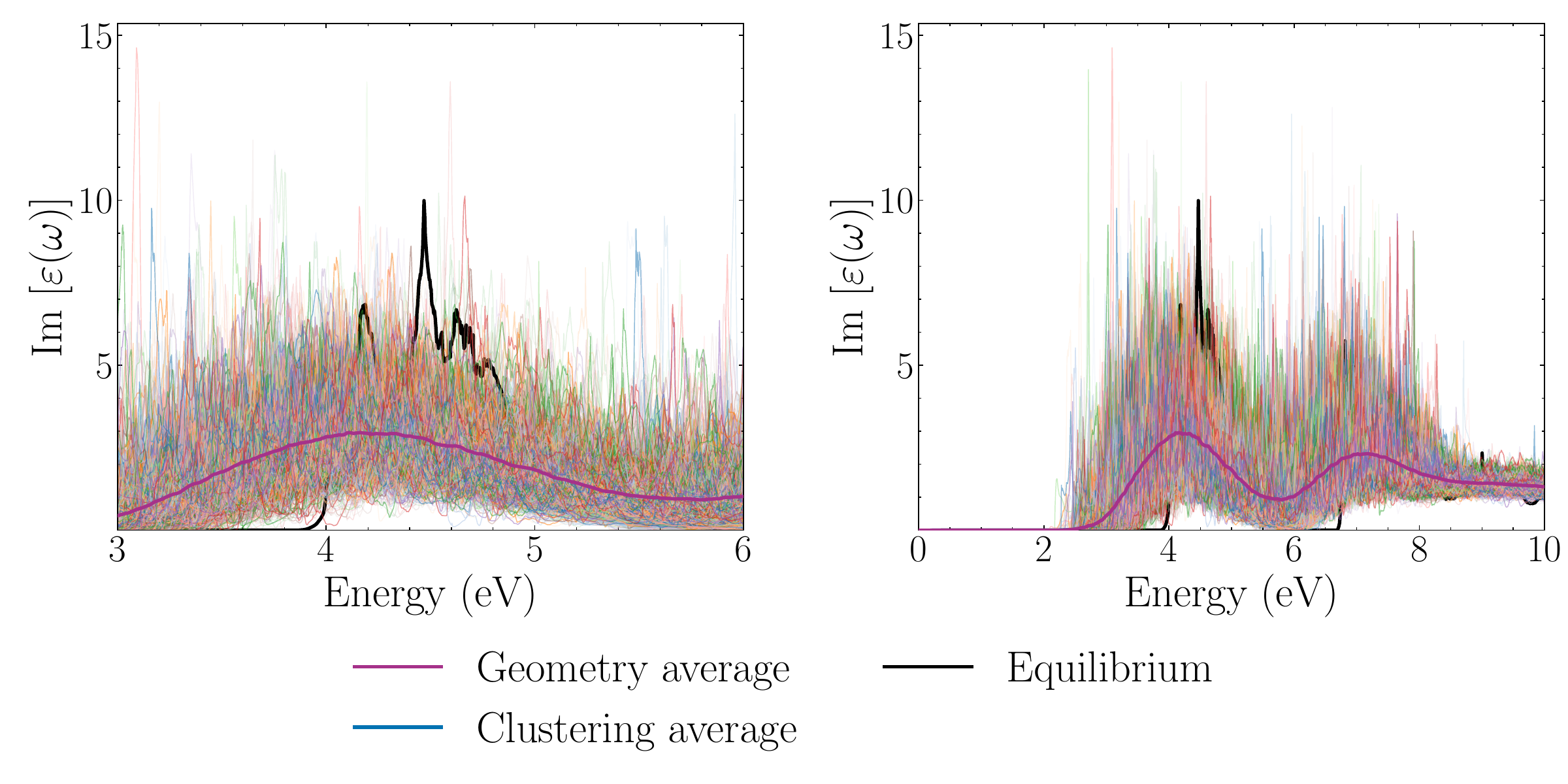}
    \caption{Imaginary part of the dielectric function of crystalline formaldehyde, $\varepsilon_2(E)$, calculated for the individual sampled distorted geometries. The left and right panels show the low-energy region and the full plotted energy range, respectively.}
    \label{fig:Form_acopvib_ind}
\end{figure}

The individual spectra exhibit pronounced variations in peak positions and intensities across the sampled geometries (Figure \ref{fig:Form_acopvib_ind}). Despite this variability, the overall shape of the ensemble-averaged spectrum is already seen in the distribution of the individual curves, particularly in the regions around 4 and 7 eV. Averaging over the geometries smooths the sharp features of the individual spectra into broader bands while retaining this common spectral profile.

\begin{figure}[H]
    \centering
    \includegraphics[width=0.98\linewidth]{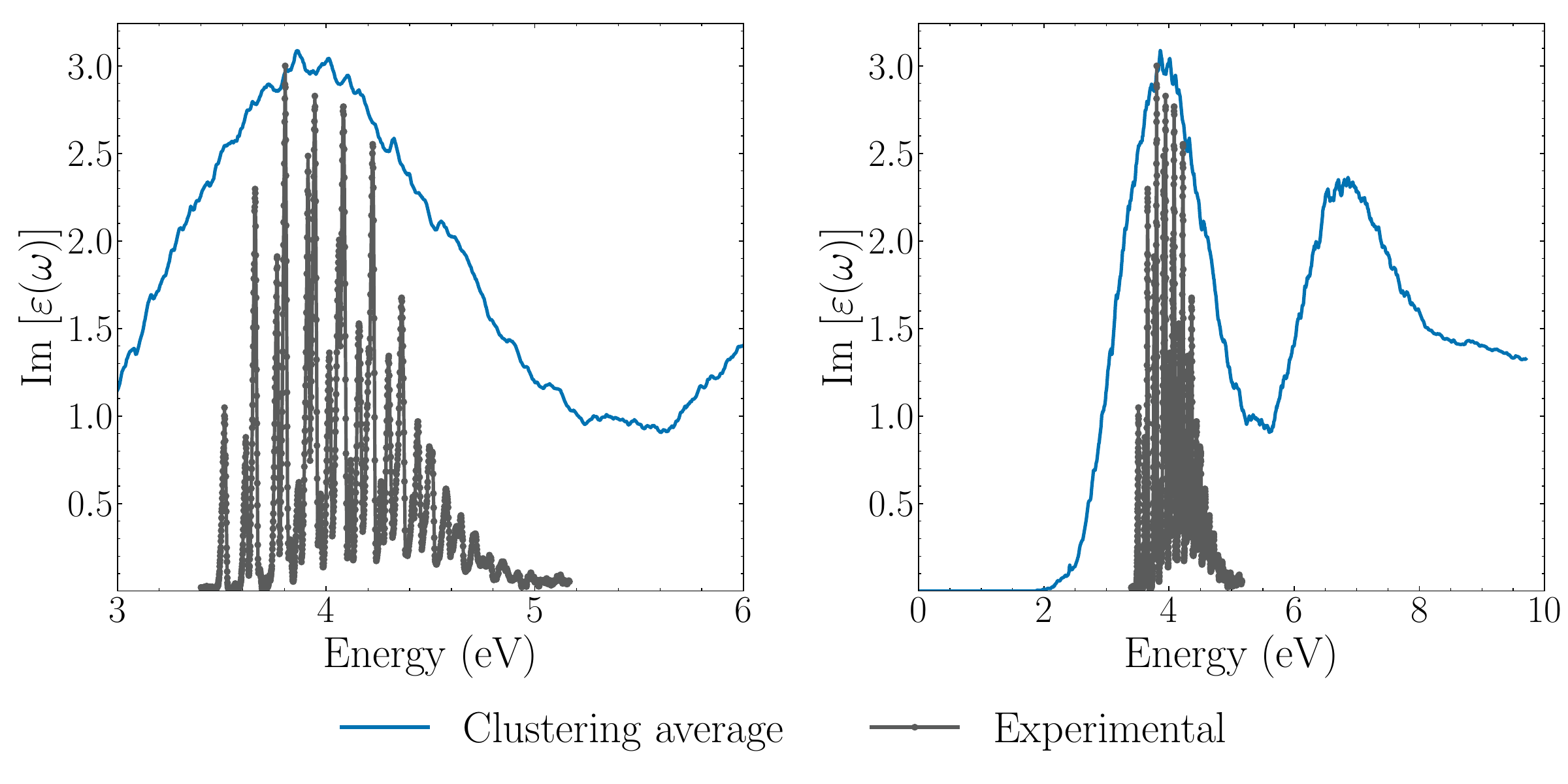}
    \caption{Comparison of the calculated spectra of crystalline formaldehyde at the equilibrium geometry and after vibrational averaging with the gas-phase absorption measurements of Gratien et al. \cite{Gratien2007}. The comparison highlights the overall envelope of the first absorption band. The calculated spectrum used for comparison with experiment was shifted by $-0.40$~eV as an empirical energy alignment.}
    \label{fig:Form_exp}
\end{figure}

The experimental spectra, obtained by Gratien et al. (2007) \cite{Gratien2007}, shows only the first band, and our calculations was able to reproduce very well the broader envelope of the experiment, as shown in Figure \ref{fig:Form_exp}.

\section{Conclusions}

We developed VCMC, a Monte Carlo protocol and Python workflow for simulating vibronic contributions to optical spectra of crystalline systems. The approach generates distorted geometries by simultaneously sampling the included harmonic phonon modes and estimates the spectrum by averaging the optical response over the resulting nuclear ensemble. To reduce the number of required electronic-structure calculations, the workflow combines structural descriptors, dimensionality reduction, and clustering to select representative geometries from the sampled ensemble.

The applications to Si, GaAs, and crystalline formaldehyde show that the method can describe vibronically induced spectral features and nuclear-motion broadening effects that are absent or poorly represented at the equilibrium geometry. For these systems, spectra obtained by averaging the optical responses of all cluster-selected geometries closely match the full-ensemble averages, showing that the selected subsets preserve the dominant spectral information with fewer electronic-structure calculations. Further benchmarks against random selection at equal calculation budgets and tests using larger supercells are needed to assess the efficiency of the selection strategy and the convergence with respect to cell size.


\begin{acknowledgement}
The authors acknowledge Conselho Nacional de Desenvolvimento e Pesquisa (CNPq), Coordenação de Aperfeiçoamento de Pessoal de N\'{\i}vel Superior (CAPES) and Fundação de Amparo \`a Pesquisa do Estado do Rio de Janeiro (FAPERJ) for financial support. Furthermore, the National Laboratory for Scientific Computing (LNCC/MCTI, Brazil) for providing HPC resources of the SDumont supercomputer, which have contributed to the research results reported within this paper (http://sdumont.lncc.br) and the DIRAC computer cluster from ”Rede de Teoria, Modelagem e Simulação de Materiais para Nanotecnologia”. MB and BCG received support from the French government under France 2030, as part of the initiative d'Excellence d'Aix-Marseille Université, Amidex (AMX-22-IN1-48, MLCHEM). The Centre de Calcul Intensif d’Aix-Marseille is acknowledged for granting access to its high-performance computing resources.
\end{acknowledgement}

\begin{suppinfo}
The Python code and all the input data used in this work can be found at https://github.com/Amanda-Dornela/VCMC.git. Tests on the number of initial geometries and the number of points per cluster are found in the supporting information.
\end{suppinfo}

\section*{Author Contribution}
Conceptualization: A.D.T., A.B.R., M.B.; Data curation: A.D.T.; Formal analysis: A.D.T.; Investigation: A.D.T.; Methodology: A.D.T., A.B.R., M.B., B.C.G., R.R.O.; Software: A.D.T., B.C.G.; Supervision: A.B.R., M.B., R.R.O.; Funding acquisition: A.B.R, R.R.O.; Project administration: M.B., A.B.R, R.R.O.; Resources: M.B., A.B.R, R.R.O.; Writing-original draft: ADT; Writing-review and editing: A.D.T., A.B.R., R.R.O., M.B., B.C.G.

\section*{Notes}
The authors declare no competing financial interest.

\bibliography{acs-template}

\newpage

\end{document}